\RequirePackage{fix-cm}
\documentclass[smallextended]{svjour3}       
\smartqed  
\usepackage{longtable}
\usepackage{graphicx}
\usepackage{lscape}
\usepackage{comment}
\usepackage[numbers]{natbib}
\usepackage{hyperref}
\begin{document}

\title{A Systematic Review on Recommender Systems in Process Mining}



\author{Mansoureh Yari Eili$^1$   \and   Jalal Rezaeenour$^{2*}$  \and Mohammadreza Fani Sani$^{3}$
}


\institute{ \at
              $^1$Department of Computer Engineering, Faculty of Technology and Engineering, University of Qom, Iran  \\
              \email{m.yari@stu.qom.ac.ir}           
           \and
           \at
              $^{2*}$Department of Industrial Engineering, Faculty of Technology and Engineering, University of Qom, Iran \\
 \email{ j.rezaee@qom.ac.ir}   
   \and
            \at
         $^3$Process and Data Science Chair, RWTH Aachen University, Aachen, Germany\\
 \email{fanisani@pads.rwth-aachen.de}   
}

\date{Received: date / Accepted: date}

\maketitle

\begin{abstract}
Considering processes of a business in a recommender system is highly advantageous. Although most studies in the business process analysis domain are of descriptive and predictive nature, the feasibility of constructing a process-aware recommender system is assessed in a few works. One reason can be the lack of knowledge on process mining potential for recommendation problems. Therefore, this paper aims to identify and analyze the published studies on process-aware recommender system techniques in business process management and process mining domain. A systematic review was conducted on 33 academic articles published between 2008 and 2020 according to several aspects. In this regard, we provide a state-of-the-art review with critical details and researchers with a better perception of which path to pursue in this field. Moreover, based on a knowledge base and holistic perspective, we discuss some research gaps and open challenges in this field. 

\end{abstract}
\section{Introduction}
\label{intro}
Most of the information systems maintain significant volumes of records of process execution events, like executing a task for a particular process instance. Process Mining (PM) is a relatively new research field   \cite{Article_1}, which intends to extract insights from the event data stored and maintained in the databases and information systems, known as event logs. PM techniques' primary objective is to apply the event data in extracting process-related information, like constructing a business process model from the observed behavior in an automated sense. An event log is a collection of events extracted in the context of a process that indicates which activity has happened at a specific time \cite{Article_2}. Moreover, a Process-Aware Information System \footnote{PAIS} is a software system able to produce, maintain and manage the event logs (e.g., ERP\footnote{Enterprise Resource Planning}  systems, CRM \footnote{Customer Relationship Management}  systems, and HISs\footnote{Hospital Information Systems}  \cite{Article_3}.

Most of the current business process analytic studies have mainly applied PM as a stand-alone application in an offline mode to discover and evaluate process models. There exist emerging efforts in predictive analytics, where typical methods mainly consist of data mining, machine learning-based prediction, and simulation \cite{Article_4}. These approaches share the following common drawbacks.

\begin{itemize}
 \item[1)] They do not suggest nor prescribe potential improvement actions to decrease the probability of inappropriate outcomes but promote the user's subjective judgment and analytical skills to induct enhancement actions.
\item[2)] The optimization procedure is executed after process complication, not during \cite{Article_5}. 
\item[3)] These techniques assume that process participants can take the most proper action manually to assure the optimal accuracy for their scenario. However, in practice, the optimal accuracy depends on different factors like different costs of process execution.
\end{itemize}

Though PM has the potential to offer many advantages to recommender systems, few studies exist where the feasibility of constructing a Process-Aware Recommender System (PARS) is assessed. Possibly operational support is the most comprehensive process mining task, where additional profits can be gained by applying the PARSs. Next to supporting the decision-making process by filling the gap between pure analysis and operational support, PARSs can be contributive in decision rules analysis adopted in different instances. For instance, offering suggestions about possible next activities or resources during the process execution, reducing the overall running time, costs, and the possible risks involved in specific situations to optimize the KPIs\footnote{Key Performance Indicators }  are some possible applications for PARSs. In general, PARSs can bring the most sizable intelligence and added value to businesses. In this context, not enough attention is directed to providing recommendations \cite{Article_6}.  

There are 28 review articles written in the PM: some have covered PM as a whole \cite{Article_7,Article_8}, some have assessed the more detailed components like algorithms comparison \cite{Article_9,Article_10,Article_11} or a specific domain, e.g., health care \cite{Article_12,Article_13,Article_14,Article_15,Article_16}, supply chain and industry \cite{Article_17,Article_18,Article_19} and education \cite{Article_20,Article_21}. There exist many review articles on process discovery \cite{Article_22,Article_23}, conformance checking \cite{Article_24,Article_25}, and predictive analytics \cite{Article_6,Article_26,Article_27,Article_28,Article_29,Article_30} in PM. As to prescriptive analytics in PM, the authors in \cite{Article_31} have assessed the process modeling recommender systems. The recommendation from the human resource point of view is evaluated by the authors in \cite{Article_32}, where 90 relevant articles on the topic of human resource allocation and recommendation applied in BPM\footnote{Business Process Management }  and PM are reviewed, with no concern in recommending the next activity. A literature review is conducted by \cite{Article_4} on prescriptive PM, where merely the proposals in their implementation sense are assessed. They classified both the predictive and prescriptive methods in machine learning, statistical analysis, and probabilistic models. 

As to the researchers here, a study is required to identify and classify the available studies regarding recommendation methods in PM in different contexts regarding the outline directions and provide researchers with a better understanding in their search in pursuing this concept. Consequently, the researchers in this field will have a holistic perspective on what has been done and the existing open challenges and future concerns. Therefore, we aim to assess the manner of recommendation techniques application in PM. 

The rest of this article is structured as follows. Sec. 2 presents preliminary concepts. Then, the research methodology is expressed in Sec. 3. Afterward, the research findings and results are provided in Sec. 4, and the discussion is run in Sec. 5. Finally, the threats to validity are expressed in Sec. 6, and Sec. 7 concludes the article. 

\section{ Preliminary concepts  }
\label{sec:1}
This section explains some fundamental concepts used in this paper, i.e.,  process mining and prescriptive business process monitoring. 
\subsection{Process mining }
\label{sec:2}
Process mining is a research discipline that provides both data-oriented and business process-oriented analysis simultaneously. There are three main sub-fields in process mining, i.e., process discovery, conformance checking, and operational support   \cite{Article_1}. 

Process discovery aims to discover a process model that accurately describes the underlying process captured within the event data. An event log is given to the process discovery algorithm, and then it produces a descriptive process model as the output that captures the control-flow information among activities in different notations such as Petri net and BPMN\footnote{Business Process Management Notation} . 

The second task is conformance checking that aims to assess to what degree a given process model and event data conform to one another. These approaches can exploit the deviations between discovered/reference process model and the corresponding event log and vice-versa to a better reflection of reality. 

Finally, the operational support aims to improve or enhance process mining results, e.g., by reflecting bottleneck information directly onto a (given) process model. In this sub-filed of process mining, we try to enrich the process model, add emphasis, provide further information (e.g., time-stamps, KPIs, providing suggestions) to enhance the process model.

\subsection{Prescriptive business process monitoring }
\label{sec:3}
The predictive business process monitoring consists of techniques applied in the upcoming status of current instances in PM \cite{Article_33} to improve the enabling proactive and corrective actions by providing on-time information. A predictive technique could predict the time necessary to execute a task in a running process instance, the next activity to be executed, or the end outcome. In the studies run in this context, suggesting or prescribing how the results can be applied in improving business processes are missing. Simultaneously, these two components should indicate how the process participants should intervene to decrease the potential of inappropriate results. To accomplish the process improvement objectives, the prediction results must be actualized in their concrete sense \cite{Article_34}.

Prescriptive business process monitoring approaches measure predictions concerning their effect on the process performance to prevent inappropriate activities therein by warning or recommending. In this study, it is revealed that as to recommendation, there exist two major methods in PM: 

\begin{itemize}
 \item[1)] \textit{Recommender systems during process modeling:} support designers in design phases \cite{Article_35,Article_36,Article_37,Article_38}. Designing business process models are highly contributive in BPM. Simultaneously, in its manual sense, it is a time-consuming and an error-prone task. However, the required time may be reduced when the models' prototypes are generated through the modeling support techniques. The recommendation methods in business process modeling assist a designer in completing the process model editing. This assistance can provide autocompleting mechanisms to choose the next best nodes for the reference process fragments to complete the workflow. The model elements' names or attachments, the beginning and ending spots, and the abstraction level of the process model can be recommended as well. These approaches can reduce the structural error count during the process design, obtain a well-defined high-quality process model, and improve the performance of the modeling procedure. 
 \item[2)]\textit{Recommender systems during the process execution:} continuously monitor process executions and predict how they are being evolved in determining the instances with a higher chance of not meeting the desired performance levels. In this way, they can provide automated recommendations of the possible best activities and resources to a running instance in order to optimize an objective function, performance, and efficiency or prevent an inappropriate outcome like case deviation, deadline, or risky execution.
\end{itemize}
 
The PARS, the focus of this study, was assessed by \cite{Article_39,Article_40} for the first time, where the process mining is applied to recommend users the possible next activity and resource, respectively. When a user requires a recommendation, first, its partial trace will be applied in the recommendation service. After that, it can be compared with the event log, and then some recommendations that would optimize the target function or match with a successfully executed behavior would be given. 

\section{Research method }
\label{sec:4}
\begin{figure}
\includegraphics[width=1\textwidth]{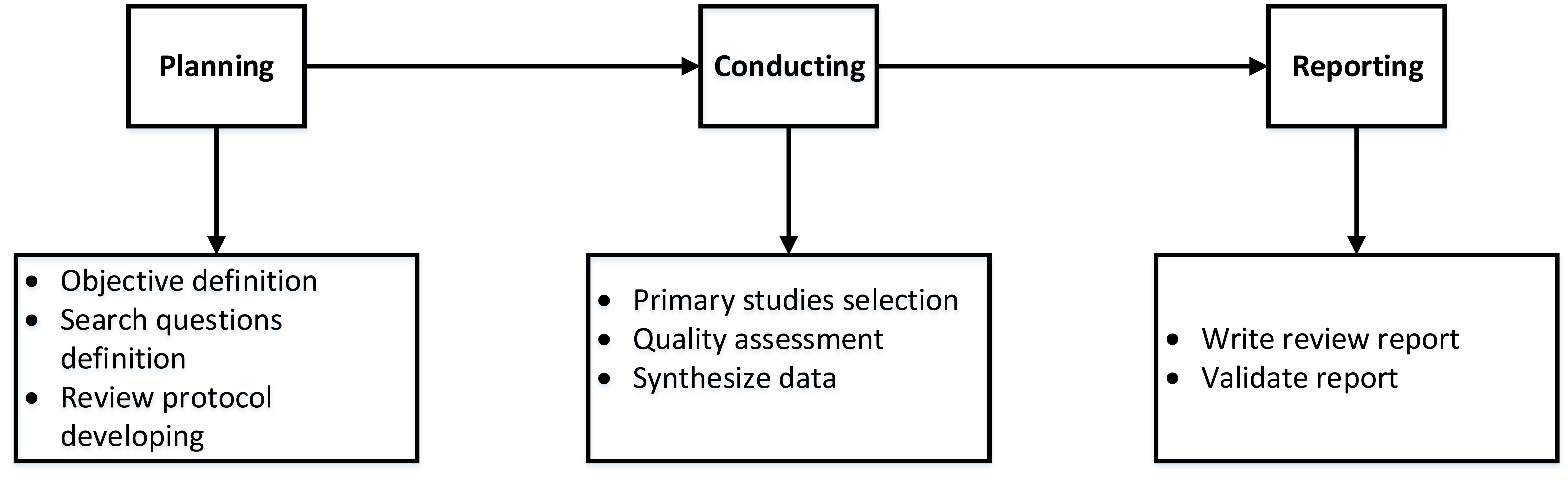}
\caption{Systematic review process \cite{Article_41}}
\label{fig1} 
\end{figure}

Here, we explain the methodology of this review article that is based on the scientific guidelines specified in \cite{Article_41} (See \autoref{fig1}), and includes the research questions, the identification of queries, and the search engines. The steps for selecting the relevant studies are provided by applying inclusion/exclusion criteria.

\subsection{Background and objectives }
\label{sec:4}
According to \cite{Article_7,Article_12}, descriptive analytics in PM has been the subject of more intense studies than other types of analytics (e.g., prescriptive). It could be a fundamental reason why the recommendation in PM has not reached the same level of maturity as other techniques \cite{Article_4}. To the best of the researchers' knowledge here, there is no extensive review study on this issue, and this is the first comprehensive review article focusing on the field of PARSs. Thus, a noble contribution in comprehending trends, challenges and filling the research gap herein. The objectives of this study are 1) to collect and characterize the studies where PM techniques are adopted to develop recommender systems, 2) refer to the existing PARSs' status, and 3) identify and classify the reviewed articles per most important aspects, like recommendation strategies, tools used, geographical areas, the application domain, where the data are acquired, and the problems are solved.

\subsection{Research questions  }
\label{sec:5}
In line with this study's objectives, the research questions are defined as to focus on specific aspects of the overall evaluation. For this purpose, we used the proposed structure in \cite{Article_42} that is presented in \autoref{tab:1}. This structure includes Population, Intervention, Comparison, Outcome, and Context (PICOC) and is proposed to extract the related concepts applicable in systemic research questions. Because this article's composition is of systematic literature review type, the comparison of interventions is not applied.

\begin{table}

\caption{PICOC structure for directing the research question \cite{Article_42} }
\label{tab:1}       
\begin{tabular}{lll}
\hline\noalign{\smallskip}
Criteria & Description  \\
\noalign{\smallskip}\hline\noalign{\smallskip}
Population & \parbox[c]{9cm}{Considering studies on how activity/resource recommendations during process execution are made in process mining}\\
Intervention & \parbox[c]{9cm}{Describe approaches (methods, strategies, techniques, and tools) applied in PARSs}   \\
Comparison & N/A\\
Outcome & Describe the effectiveness of PARSs\\
Context & \parbox[c]{9cm}{Describe the usage domain: primary studies in the BPM and PM disciplines} \\
\noalign{\smallskip}\hline
\end{tabular}
\end{table}

In this paper, the research focuses are on the next activity/resource recommendations during process execution. The research questions mainly focus on discovering, identifying, and evaluating the approaches intended to apply recommendation techniques in the PM context. Here, we aim to answer the following research questions.

\begin{itemize}
 \item[1)] What are the trends and statistic of case studies where the PM techniques and algorithms were applied in the recommendation domain between 2008 and 2020? 
\end{itemize}
This question aims to determine the rise and fall of the trend in published studies where PARSs are addressed in the given period. It is possible to identify the type of venues where the relevant studies are being published, contributing to the perception of relevant active journals and conferences in this field. 
\begin{itemize}
 \item[2)]What are the common characteristics of the studies in the field of PARS?
\end{itemize}

For answering this question, the reviewed articles are classified according to different aspects, i.e., recommendation methodologies and approaches, applied tools, type of implementation, evaluation methods, geographical information, data type and application domains.   

\subsection{Identification of search engines, keywords, and queries  }
\label{sec:6}
In the third step of conducting this literature review, the search queries expressed in \autoref{tab:2} are applied in the keywords, title, or abstract of published articles between 2008 and 2020. We investigated several popular digital libraries, i.e., ACM, Google Scholar, Emerald, IEEE Explore, Pubmed, Science Direct, and Springer Link. These search queries are the commonly-applied keywords in the relevant articles on PARSs.

\begin{table}
\caption{Generic Search Queries }
\label{tab:2}       
\begin{tabular}{lll}
	
\noalign{\smallskip}\hline\noalign{\smallskip}
\multicolumn{1}{c}{'Recommender system' OR 'Decision support' OR 'Prescriptive analytics' OR 'Next}  \\
\multicolumn{1}{c}{recommendation' OR 'Next action recommendation' OR 'Next resource }\\
\multicolumn{1}{c}{ recommendation' AND 'Process mining' }\\
\noalign{\smallskip}\hline
\end{tabular}
\end{table}

The results obtained from each digital library are tabulated in \autoref{tab:3}. A total of 2253 primary studies are obtained through the \textit{Identifying all relevant studies} phase in September 2020. After running the digital library search, all irrelevant content (1035 articles) and duplicates (72 articles) are excluded for the next phase. Thus, 1146 articles remained for further screening. This literature review is conducted subject to the inclusion and exclusion criteria, which are explained in the following subsection to assure unbiased selection.

\begin{table}
\caption{Published articles in journals, Conferences and PhD./Master thesis}
\begin{center}
\label{tab:3}       
\begin{tabular}{lll}
\hline\noalign{\smallskip}
Digital Library & Search Results  \\
\noalign{\smallskip}\hline\noalign{\smallskip}
\multicolumn{1}{c}{ACM} &\multicolumn{1}{c}{ 106 } \\
\multicolumn{1}{c}{Google Scholar} &\multicolumn{1}{c} {1395}  \\
\multicolumn{1}{c}{Emerald} & \multicolumn{1}{c}{111}  \\
\multicolumn{1}{c}{IEEE Explore} &\multicolumn{1}{c}{ 71}  \\
\multicolumn{1}{c}{Pubmed} & \multicolumn{1}{c}{56}  \\
\multicolumn{1}{c}{Science Direct} &\multicolumn{1}{c}{ 461}  \\
\multicolumn{1}{c}{Springer} &\multicolumn{1}{c}{ 53}  \\
\multicolumn{1}{c}{Total} & \multicolumn{1}{c}{2253}  \\
\noalign{\smallskip}\hline
\end{tabular}
\end{center}
\end{table}

\subsection{Inclusion and exclusion criteria }
\label{sec:7}
In this phase, the inclusion criteria are applied to add a study to the analysis, and exclusion criteria are applied to exclude one from the analysis. 

The inclusion criteria are as follows:

\begin{itemize}
\item[-]IC-1: The recommendation method/algorithm/framework for the next activity/resource recommendation within the process mining domain.
\item[-]IC-2: Peer-reviewed articles in scientific journals, conferences, or workshops.
\item[-]IC-3: The article is written in English
\end{itemize}

The exclusion criteria are as follows:
\begin{itemize}
\item[-]EC-1: The relation is only to one of the two areas considered in process mining or recommendation areas. 
\item[-]EC-2: Concerning the process modeling recommender systems. 
\item[-]EC-3: The focus is on resource allocation as to process performance improvement but not based on improving decision-making through process mining.
\item[-]EC-4: The article does not include a method/experiment/case study that is implemented and evaluated.
\end{itemize}

Based on these criteria, a total of 105 studies are considered in this study. Because this phase consists of studies with no contribution to the next activity/resource recommendation, the \textit{screening of papers} phase is run to skim the metadata (i.e., title, abstract, and keywords sections) or those that do not correspond with IC1. After the screening phase, a total of 52 articles were extracted and remained for the \textit{full reading of the articles} phase. After full reading, some articles were omitted according to EC-2 and EC-4, and reduced the count to a final set of 33 studies (13 journal papers, 19 conference papers, and four postgraduate and doctoral dissertations which were not on the scope of our review). The results obtained after applying each phase during the selection process are tabulated in \autoref{tab:4}.

\begin{table}
\caption{Results obtained after applying each phase}
\label{tab:4}       
\begin{tabular}{lll}
\hline\noalign{\smallskip}
\multicolumn{1}{c}{Phase} &\multicolumn{1}{c}{ Amount of papers} \\
\noalign{\smallskip}\hline\noalign{\smallskip}
\multicolumn{1}{c}{Identifying all relevant studies} & \multicolumn{1}{c}{2253}\\
\multicolumn{1}{c}{Removing duplicate or irrelevant studies} & \multicolumn{1}{c}{1146}\\
\multicolumn{1}{c}{Applying the inclusion and exclusion criteria} & \multicolumn{1}{c}{105}\\
\multicolumn{1}{c}{Screening of papers} & \multicolumn{1}{c}{52}\\
\multicolumn{1}{c}{Full reading} &\multicolumn{1}{c}{ 33}\\
\noalign{\smallskip}\hline
\end{tabular}
\end{table}

\section{Results }
\label{sec:8}
\begin{figure}
	
\includegraphics[width=0.7\textwidth]{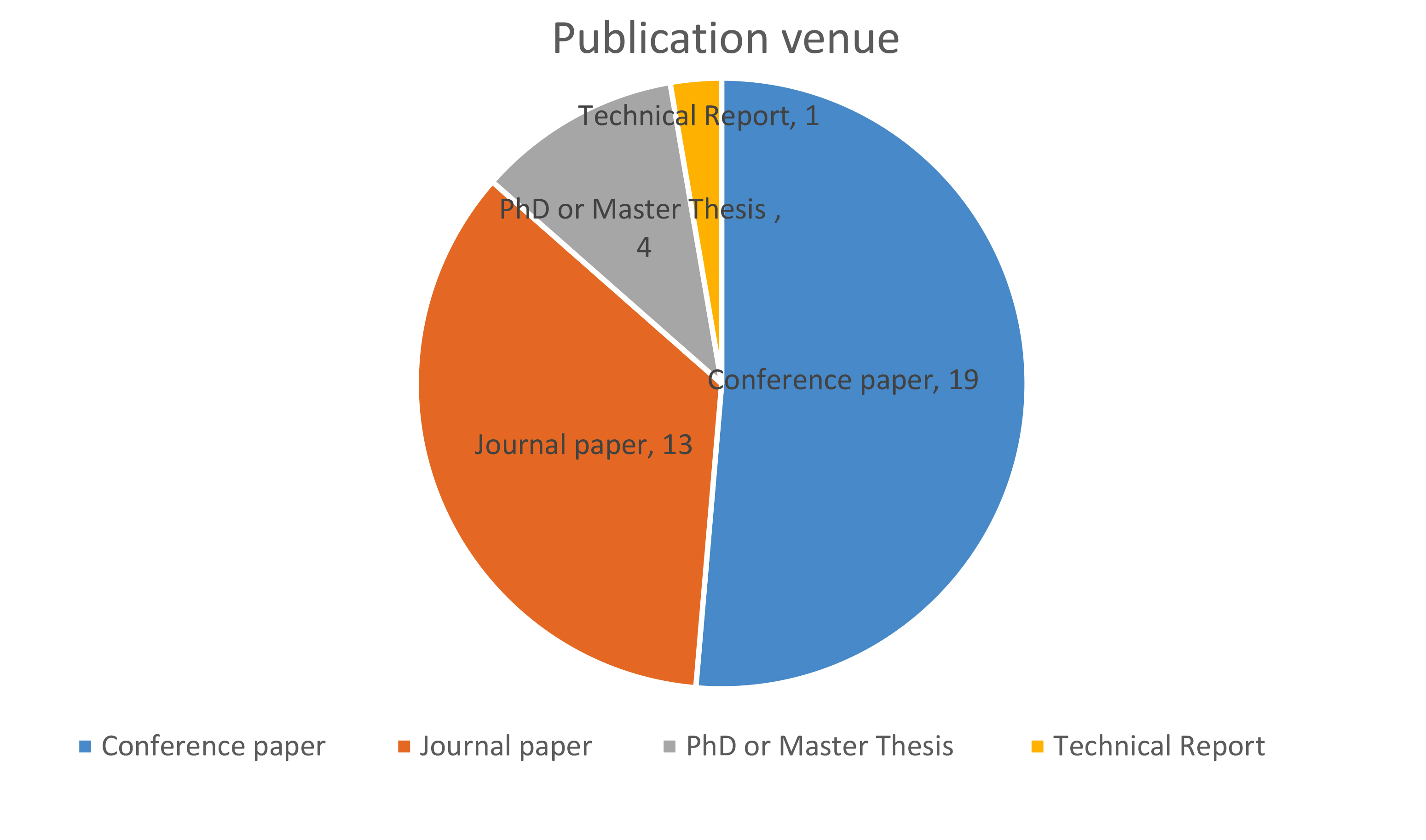}
\caption{Studies about PARS by venue}
\label{fig2}
\end{figure}

This section represents the undertaken data extraction phase results according to the protocol shown in the previous section. \autoref{fig2} shows the publication venues of the selected research articles. Results show that the articles that belong to indexed journals make 35\% of all. The relevant articles presented in conferences make up 51\% of the total, and the International Conference on Business Process Management was make up for publishing seven articles of 19 conference articles. Moreover, 3\% of articles belong to the technical report category, and 11\% corresponds to postgraduate and doctoral theses that were not on our review's scope.
 
\begin{figure}
\includegraphics[width=1.1\textwidth]{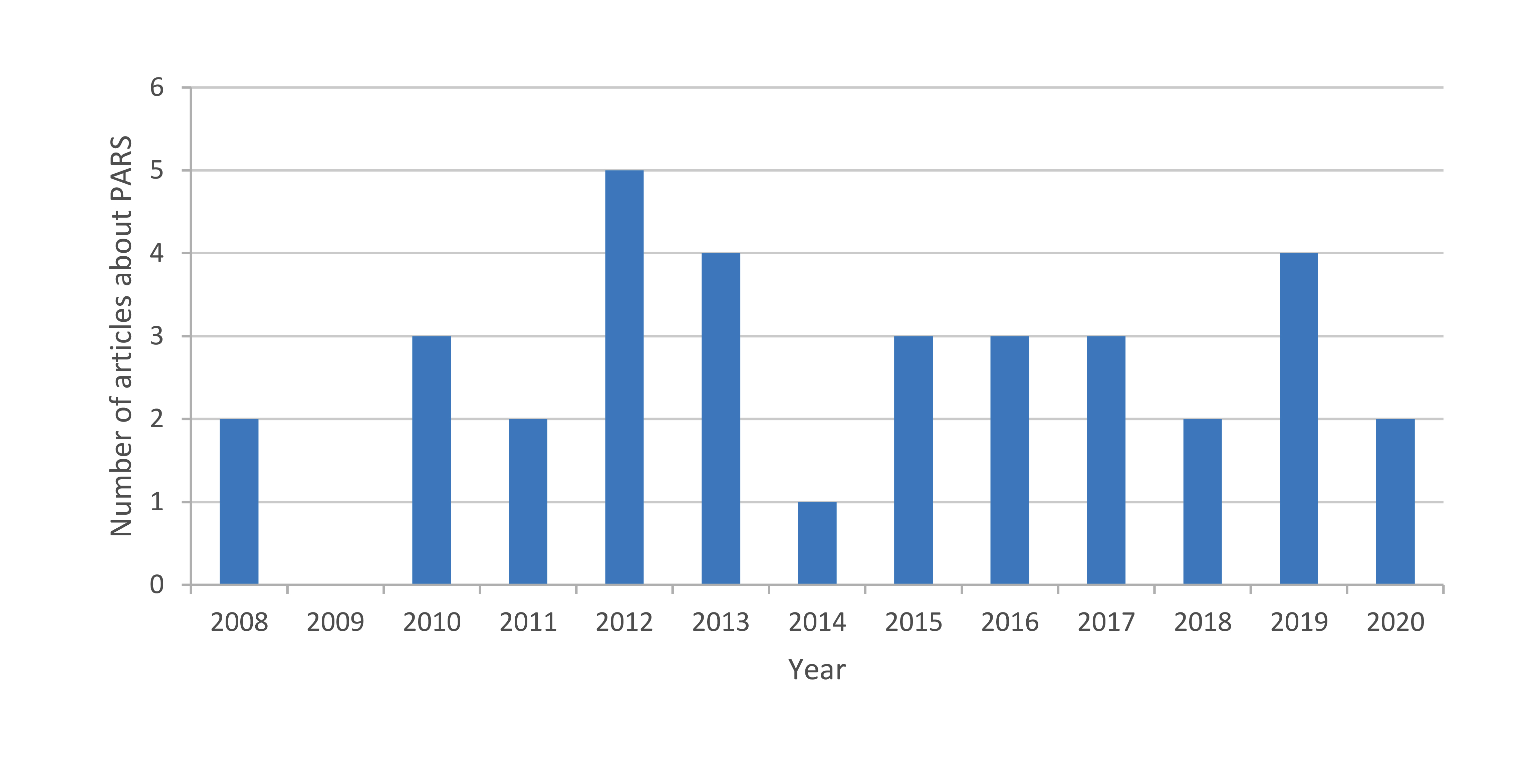}
\caption{Annual published articles about PARSs (2008-2020)}
\label{fig3}
\end{figure}

Concerning the distribution of studies about PARSs during the past 12 years, \autoref{fig3}, more than half of these studies are disseminated during the past five years, indicating a recent surge of interest in operational support tasks in PM. A complete list of these studies, including the authors' name, title, publication date, and method(s) is tabulated in \autoref{tab:5}.

\begin{table}[p]
	
	\caption{Classification of researches regarding PARSs}
	\label{tab:5}
	\tiny
	\begin{tabular}{|p{2cm}|p{2.282cm}|p{2cm}|p{1.45cm}|p{1.0cm}|p{2.5cm}|p{0.5cm}|p{1.0cm}|p{1.0cm}|}
		\hline
		\textbf{Authors}                             & \textbf{Publication}                                                               & \textbf{Recommendation Type}                  & \textbf{Process Perspectives}        & \textbf{Contextual Information} & \textbf{Application Domain and Data Type}                        & \textbf{Tools}      & \textbf{Evaluation}  & \textbf{Location}            \\ \hline
		Schonenberg,   et. al., (2008) {[}39{]}      & International Conference on BPM                                                    & NAR, Metric-based                             & Control flow                         & -                               & -                                                                & ProM                & Proposal of Solution & the Netherlands, Austria          \\ \hline
		Liu, et.   al., (2008) {[}40{]}              & Computers in Industry                                                              & NRR- Pattern optimizing                       & Control flow                         & -                               & Manufacturing enterprises                                        & WEKA                & Evaluation Research  & China Australia              \\ \hline
		Dorn,   et. al., (2010) {[}47{]}             & International Conference on  BPM                                                   & NAR, Metric-based                             & Control flow                         & Case level                      & web usage data                                                   & web-based prototype & Evaluation Research  & Austria, Germany             \\ \hline
		van der   Aalst, et. al., (2010) {[}44{]}    & International Conference on   Advanced Information Systems Engineering             & NAR, Metric-based                             & Time                                 & -                               & -                                                                & ProM                & Validation Research  & the Netherlands                   \\ \hline
		Motahari-Nezhad,   et. al.,  (2011) {[}72{]} & International Conference on  BPM                                                   & NAR/NRR,  Pattern optimizing and Metric-based & Control flow and resource            & Case level                      & Human resource                                                   & Eclips              & Evaluation Research  & USA                          \\ \hline
		Haisjackl,   et. al., (2010) {[}43{]}        & International Conference on  BPM                                                   & NAR, Metric-based                             & Control flow                         & -                               & -                                                                & ProM                & Proposal of Solution & Austria                      \\ \hline
		Huang, et.al.,   (2011) {[}61{]}             & Expert Systems with Applications                                                   & NRR, Pattern optimizing                       & Resource                             & -                               & healthcare                                                       & Weka                & Validation Research  & China                        \\ \hline
		Chen, et.al., (2013)                         & Personal and ubiquitous   computing                                                & NAR, Pattern optimizing                       & Control flow                         & Case level                      & Web-based application                                            & -                   & Validation Research  & Japan                        \\ \hline
		Schönig, et.al.,   (2012) {[}70{]}           & International Conference on   Collaborative Computing                              & NRR, Pattern optimizing                       & Data and Resource                    & Case level                      & -                                                                & Weka                & Validation Research  & Germany                      \\ \hline
		Petrusel, et.al.,   (2012) {[}51{]}          & In International Conference   on Business Information Systems                      & NAR, Metric-based                             & Control-flow, Data and Resource      & -                               & -                                                                & ProM                & Validation Research  & Romania                      \\ \hline
		Liu, et.al., (2012),   {[}68{]}              & Knowledge-Based Systems                                                            & NRR, Pattern optimizing                       & Control flow and Resource            & -                               & Manufacturing                                                    & Weka                & Validation Research  & China                        \\ \hline
		Barba, et.al.,   (2011) {[}45{]}             & International Conference on  BPM                                                   & NAR/ NRR, Metric based                        & Control-flow and Resource            & -                               & -                                                                & ConDec              & Proposal of Solution & Austria                      \\ \hline
		Nakatumba, et.al.,   (2012) {[}50{]}         & BPM center report BPM-12-05                                                        & NAR, Pattern optimizing                       & Control-flow                         & -                               & -                                                                & ProM                & Proposal of Solution & the Netherlands                   \\ \hline
		Khodabandelou,   et.al., (2013) {[}78{]}     & International Conference on   Research Challenges in Information Science           & NAR                                           & -                                    & -                               & Human resource                                                   & Eclips, ProM        & Validation Research  & France                       \\ \hline
		Triki, et.al.,   (2013) {[}59{]}             & Workshops on Enabling   Technologies: Infrastructure for Collaborative Enterprises & NAR, Pattern optimizing                       & Control flow                         & -                               & -                                                                & Web-based prototype & Proposal of Solution & Tunisia, France              \\ \hline
		Obregon, et.al.,   (2013), {[}67{]}          & In Asia-Pacific Conference   on  BPM                                               & NRR, Metric-based and Pattern   optimizing    & Time and Resource                    & -                               & Financial , Sales/ marketing                                     & ProM                & Validation Research  & South Korea                  \\ \hline
		Conforti,   et.al., (2015) {[}66{]}          & Decision Support Systems                                                           & NRR, Risk minimizing                          & Control-flow, Time, Resource         & Case level                      & Insurance                                                        & YAWL, WEKA          & Proposal of Solution & Australia, the Netherlands, Italy \\ \hline
		Gröger, et.al.,   (2014) {[}48{]}            & International Conference on  BPM                                                   & NAR, Pattern optimizing                       & Control-flow and Resource            & Process level                   & Manufacturing                                                    & RapidMiner          & Validation Research  & Germany                      \\ \hline
		Zhao, et.al., (2015)   {[}71{]}              & International Conference on   Intelligent Computing                                & NRR, Metric-based                             & Resource                             & -                               & -                                                                & -                   & Validation Research  & China                        \\ \hline
		Huber, et.al.,   (2015) {[}49{]}             & International Conference on   Subject-Oriented  BPM                                & NAR, Metric-based and Pattern   optimizing    & Control-flow and Time                & -                               & Transport                                                        & Weka,  CoCaMa       & Proposal of Solution & Germany                      \\ \hline
		Zhao, et.al., (2016)   {[}73{]}              & Knowledge and Information   Systems                                                & NRR, Metric-based                             & Resource                             & Case level                      & Healthcare                                                       & -                   & Validation Research  & China                        \\ \hline
		Arias, et.al.,   (2016) {[}65{]}             & International Conference on  BPM                                                   & NRR, Metric-based                             & Resource                             & Process level                   & Sales/ marketing                                                 & ProM                & Validation Research  & Chile                        \\ \hline
		Sindhgatta, et.al.,   (2016) {[}75{]}        & International Conference on   Advanced Information Systems Engineering             & NRR, Pattern optimizing                       & Control-flow and Resource            & Process level                   & Car repair and maintenance Volvo   IT Belgium process (BPIC 2013 & -                   & Validation Research  & India,  Australia            \\ \hline
		Zhao, et.al.,  (2017) {[}57{]}               & Information Systems Frontiers                                                      & NRR, Metric-based                             & Resource                             & -                               & Sales/ marketing                                                 & -                   & Validation Research  & China                        \\ \hline
		Yang, et.al.,   (2017){[}60{]}               & ACM International Conference on   Knowledge Discovery and Data Mining              & NAR, Metric-based and Pattern   optimizing    & Control flow                         & Process level                   & Healthcare                                                       & Java                & Evaluation Research  & USA                          \\ \hline
		Schobel et.al.,   (2017) {[}52{]}            & In Advances in Intelligent   Process-Aware Information Systems                     & NRR, Metric-based                             & Control flow                         & -                               & E-commerce                                                       & Matlab              & Evaluation Research  & Germany                      \\ \hline
		Terragni, et.al.,   (2018) {[}54{]}          & International Conference on   Future Internet of Things and Cloud                  & NRR, Metric-based                             & Control flow                         & -                               & Web-based application                                            & Disco               & Evaluation Research  & the Netherlands                   \\ \hline
		Seeliger, et.al.,   (2018) {[}53{]}          & Workshop on Interactive   Data Exploration and Analytics                           & NAR, Pattern optimizing                       & Control-flow, Time, Data,   resource & -                               & Sales/ marketing                                                 & ProM                & Validation Research  & Germany                      \\ \hline
		Pereira Detro   et al., (2020) {[}58{]}      & Enterprise Information Systems                                                     & NAR, Pattern optimizing                       & Control flow                         & -                               & Healthcare                                                       & ProM                & Evaluation Research  & Brazil                       \\ \hline
		Park, et.al.,   (2019){[}69{]}               & International Conference on   Process Mining                                       & NRR, Metric-based                             & Control-flow and Resource            & -                               & Financial                                                        & -                   & Validation Research  & South Korea                  \\ \hline
		Dees, et.al.,   (2019)  {[}74{]}             & International Conference on  BPM                                                   & NAR, Risk minimizing                          & Control flow                         & -                               & Insurance                                                        & Sci-kit learn       & Validation Research  & Germany                      \\ \hline
		Weinzierl,   et.al., (2020) {[}55{]}         & International Conference on  BPM                                                   & NAR                                           & Control flow                         & -                               & Sales/ marketing                                                 & Tensor flow         & Validation Research  & Germany                      \\ \hline
		Weinzierl,   et.al., (2020) {[}56{]}         & Hawaiian International   Conference on System Sciences                             & NAR, Metric-based                             & Control flow                         & -                               & Web-based application                                            & Tensor flow         & Validation Research  & Germany                      \\ \hline
		Ghazaleh   Khodabandelou (2013) {[}79{]}     & PhD Thesis                                                                         & \multicolumn{7}{l|}{Université   Panthéon-Sorbonne - Paris}                                                                                                                                                                                                           \\ \hline
		Irene Teinemaa   (2019)                      & PhD Thesis                                                                         & \multicolumn{7}{l|}{University of TARTU,   Estonia}                                                                                                                                                                                                                   \\ \hline
		Sen Yang (2019)                              & PhD Thesis                                                                         & \multicolumn{7}{l|}{Rutgers University, Texas USA}                                                                                                                                                                                                                    \\ \hline
		Alessandro   Terragni (2018)                 & Master Thesis                                                                      & \multicolumn{7}{l|}{Eindhoven university   of technology, the Netherlands}                                                                                                                                                                                                \\ \hline
	\end{tabular}
\end{table}

\subsection{Recommendation approaches and strategies }
\label{sec:8}
The available techniques for PARSs are classified according to \textit{recommendation approaches} and \textit{recommendation strategies}. In recommendation approaches, the focus is on whether the approach considers the control-flow perspective (next activity recommendation) or the organizational perspective (next resource recommendation) to make a recommendation. Moreover, recommendation strategies consider whether the recommendation production strategy is of pattern optimization, risk minimization, or metric-based.

\subsubsection{Next Activity Recommendation (NAR)}
\label{sec:9}
By having a particular partial trace that reflects the process instances' current state, users can select which activity should be executed next. This selection is challenging because process performance objectives (e.g., minimization of running time) should be of concern. It is evident that users with little experience have more difficulties during the process execution, and this may lead to having unfinished process instances \cite{Article_43}. In this context, when the users are unfamiliar, recommendation support is essential during process execution.

NAR methods, according to the type of recommended element, can be divided into (1) single activity recommendation \cite{Article_43,Article_44,Article_45,Article_46,Article_47,Article_48,Article_49,Article_50,Article_51,Article_52,Article_53,Article_54,Article_55,Article_56,Article_57,Article_39} and (2) multi-activity recommendation, named trace or path recommendation as well, initially recommend all the possible steps at once \cite{Article_58,Article_59,Article_60}. The path recommendation problem in the multi-process instance environment is yet to be addressed. 

\subsubsection{Next Resource Recommendation (NRR)}
\label{sec:10}
The NAR techniques merely consider the control-flow perspective and provide recommendations about the next possible steps, with no idea about the next resources. The resource is an essential index in business process performance. Concerning the application area, resources can be equipment, human resources, money, or software systems that perform the business process activities \cite{Article_61}.

Resource allocation, known as staff assignment, is considered necessary in BPM. The proper allocation of the most appropriate activity would lead to effective process performance and efficiency, service level, resource utilization, and reduce execution costs \cite{Article_32}. There are many resource assignment mechanisms in the BPM literature \cite{Article_62,Article_63,Article_64}. Still, focusing on actively recommending resources by applying the workflow information is rare.

 NRR methods analyze resource-related data, manage execution activities of a business process and assure that the right resources are assigned to the right tasks at the right time. Based on the application environment, NRRs can be divided into single instance resource recommendation \cite{Article_40,Article_61,Article_65,Article_66,Article_67,Article_68,Article_69,Article_70,Article_71} and multi-instance resource recommendation. However, the resource recommendation problem in the multi-process instance environment is still not well solved. 

The two research approaches proposed by \cite{Article_45,Article_72} cover both the NAR and NRR approaches. In these studies, both the best next activities in handling a new process instance and recommendations on resources to assist in the resolution of the instance, are applied.

\subsubsection{Recommendation strategy }
\label{sec:11}
Based on the reviewed articles in the PARS domain, another classification is compiled according to the strategies applied for providing recommendations. The first recommendation strategy is the \textit{metric-based strategy}, where goal-supporting activities or resources are recommended as the pre-defined goal optimization (e.g., reducing cost or cycle time). In most studies run on NAR approaches \cite{Article_39,Article_43,Article_44,Article_45,Article_47,Article_51,Article_52,Article_54,Article_55,Article_56} and in some of NRR approaches \cite{Article_57,Article_65,Article_69,Article_71,Article_73}, this strategy is applied.
 
The second recommendation strategy is the \textit{risk-minimizing} strategy, focusing on risks concerning metric deviations formulated as a mathematical problem. Such studies focus on potential risk reduction during process execution and support users in making risk-informed decisions. If the process instance is executed in such a manner, the likelihood and severity of an error occurrence are calculated. An alarm may be activated in a running instance to intervene, mitigate or inhibit the improper outcomes and reduce cost. Only two articles study this issue \cite{Article_66,Article_74}. 

The third recommendation strategy is \textit{pattern optimization strategy} which includes pattern optimization, statistical and data mining techniques. This strategy is implemented in two steps, i.e., discovering traces of executed instances and then matching them with a running instance. This strategy is applied in six studies in NAR \cite{Article_46,Article_48,Article_50,Article_53,Article_58,Article_59} and five studies in NRR \cite{Article_40,Article_61,Article_68,Article_70,Article_75} approaches.   

\begin{figure}
	
\includegraphics[width=1\textwidth]{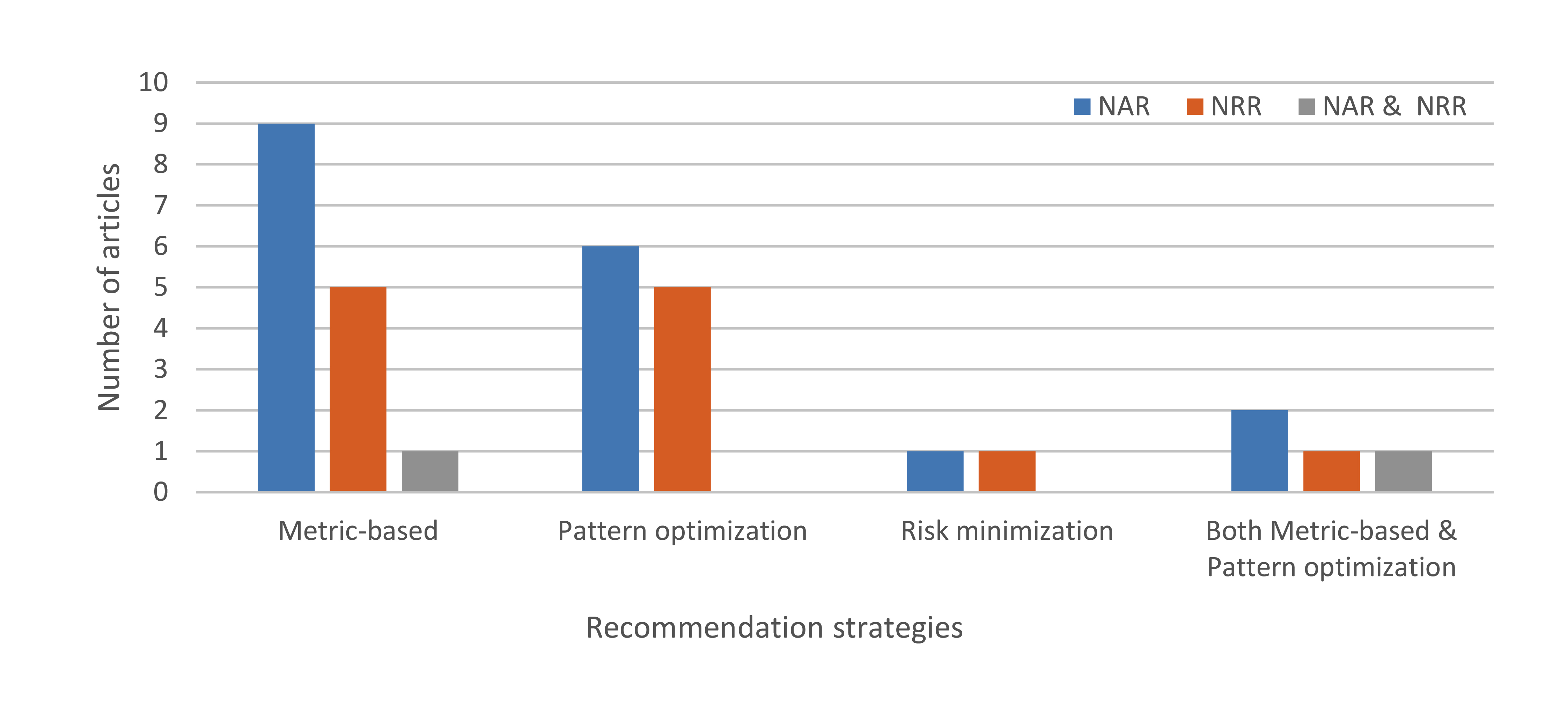}
\caption{ARecommendation approaches and strategies about PARSs}
\label{fig4}
\end{figure}

Both the \textit{metric-based} and \textit{pattern optimization strategies} are addressed in four articles \cite{Article_49,Article_60,Article_67,Article_72}. The classifications of all studies based on recommendation strategies and approaches are shown in \autoref{fig4}. 

\subsection{Process perspective and contextual information   }
\label{sec:12}
	The process perspectives are divided into the following four categories. The focus of the \textit{Control-flow perspective}, is on the process activities order, and applied in 66\% \cite{Article_39,Article_43,Article_45,Article_46,Article_47,Article_48,Article_49,Article_50,Article_51,Article_52,Article_53,Article_54,Article_56,Article_58,Article_59,Article_60,Article_66,Article_68,Article_69,Article_72,Article_74,Article_75}. The \textit{time perspective} focuses on the execution time and frequency of activities, analyzing bottlenecks, waiting time, and predict the remaining time of running instances, applied in 15\% \cite{Article_44,Article_49,Article_53,Article_67,Article_66}. The features of instances that constitute the \textit{data perspective}, is applied in 10\% \cite{Article_51,Article_53,Article_70}. The \textit{organizational perspective}, focuses on the collaboration and relation between resources, is applied in 47\% \cite{Article_40,Article_45,Article_48,Article_51,Article_53,Article_57,Article_61,Article_65,Article_66,Article_67,Article_68,Article_69,Article_70,Article_71,Article_72,Article_73,Article_75}. 

All perspectives are considered by the authors in \cite{Article_53}; the control-flow, resource, and time perspectives are considered in  \cite{Article_66}; while the control-flow, resource, and data perspectives are considered in \cite{Article_51}. Moreover, control-flow and resource are reflected in \cite{Article_45,Article_48,Article_68,Article_69,Article_72,Article_75}.

The process mining should be contextualized \cite{Article_53,Article_76}. It means that it is required to extend the scope of PM and extract more insights from the event log. Context is defined in \cite{Article_77} as \textit{any information reflecting the changing circumstances during a business process's execution}. Based on the literature review made here, studies usually consider the instance and activity attributes. Some of the PARSs only focus on the instance level \cite{Article_46,Article_48,Article_60,Article_66,Article_70,Article_73,Article_78} and just a few on process level information \cite{Article_65,Article_75}. Recently, a meta-model is defined based on a process cube, which consists of context entities and context attributes \cite{Article_65}. Context entities are connected through the context relationships, though none distinguishes the impact of external factors in the process. 

\subsection{Data type and application domain  }
\label{sec:13}
The reviewed articles can be classified according to the type of applied data. The methods are assessed through the three types of event logs: (1) real-life logs (i.e., extracted from real-life process execution), (2) synthetic logs (i.e., generated through simulation or replaying a real-life model), and (3) artificial logs (i.e., generated by replaying an artificial model). It is revealed that most methods, 24 out of 33 are tested through real-life logs \cite{Article_5,Article_39,Article_48,Article_49,Article_51,Article_52,Article_53,Article_54,Article_56,Article_57,Article_58,Article_60,Article_61,Article_65,Article_66,Article_67,Article_68,Article_69,Article_70,Article_71,Article_73,Article_74,Article_75,Article_78}. Among them, two approaches in \cite{Article_67,Article_75} are further tested against synthetic logs, while one \cite{Article_69} is tested against artificial logs. Four approaches are tested on synthetic logs \cite{Article_44,Article_46,Article_47,Article_72}, while five articles are tested on artificial logs only \cite{Article_39,Article_43,Article_45,Article_59,Article_70}. Nine articles have applied the publicly-available logs instead of the private logs, which prevent the replicability of the results because they are not accessible. The details on these event logs are expressed in Appendix.

PARSs have been applied in application domains to address varying problems. As to the application domains in the real-life logs, it is found that a set of the logs introduced by the Business Process Intelligence Challenge (BPIC) are applied by 6 studies. The BPIC, as part of the BPM conference series, is held on an annual basis. These logs are publicly available through the 4TU.center repository of Eindhoven University and cover domains like purchase order handling process (BPIC 2019) applied in \cite{Article_55}, Employee Insurance Agency (BPIC 2016), applied in  \cite{Article_66}, Dutch financial institute (BPIC 2012, 2017), applied in \cite{Article_67,Article_69}, and Volvo IT Belgium process (BPIC 2013),  \cite{Article_75}. 

Next to these publicly available real-life logs, a range of private real-life logs originating from different domains like healthcare \cite{Article_58,Article_60,Article_61,Article_73}, insurance company \cite{Article_66}, manufacturing \cite{Article_48,Article_57,Article_68}, web usage data \cite{Article_46,Article_47,Article_53,Article_54,Article_56}, IT service continuity management \cite{Article_72}, ACM system \cite{Article_49} and helpdesk process \cite{Article_55,Article_65} are applied as well. Publicly available synthetic event logs like repairing telephone, car repair, and maintenance are applied in \cite{Article_67} and \cite{Article_75}, respectively.

\begin{figure}
	
\includegraphics[width=1\textwidth]{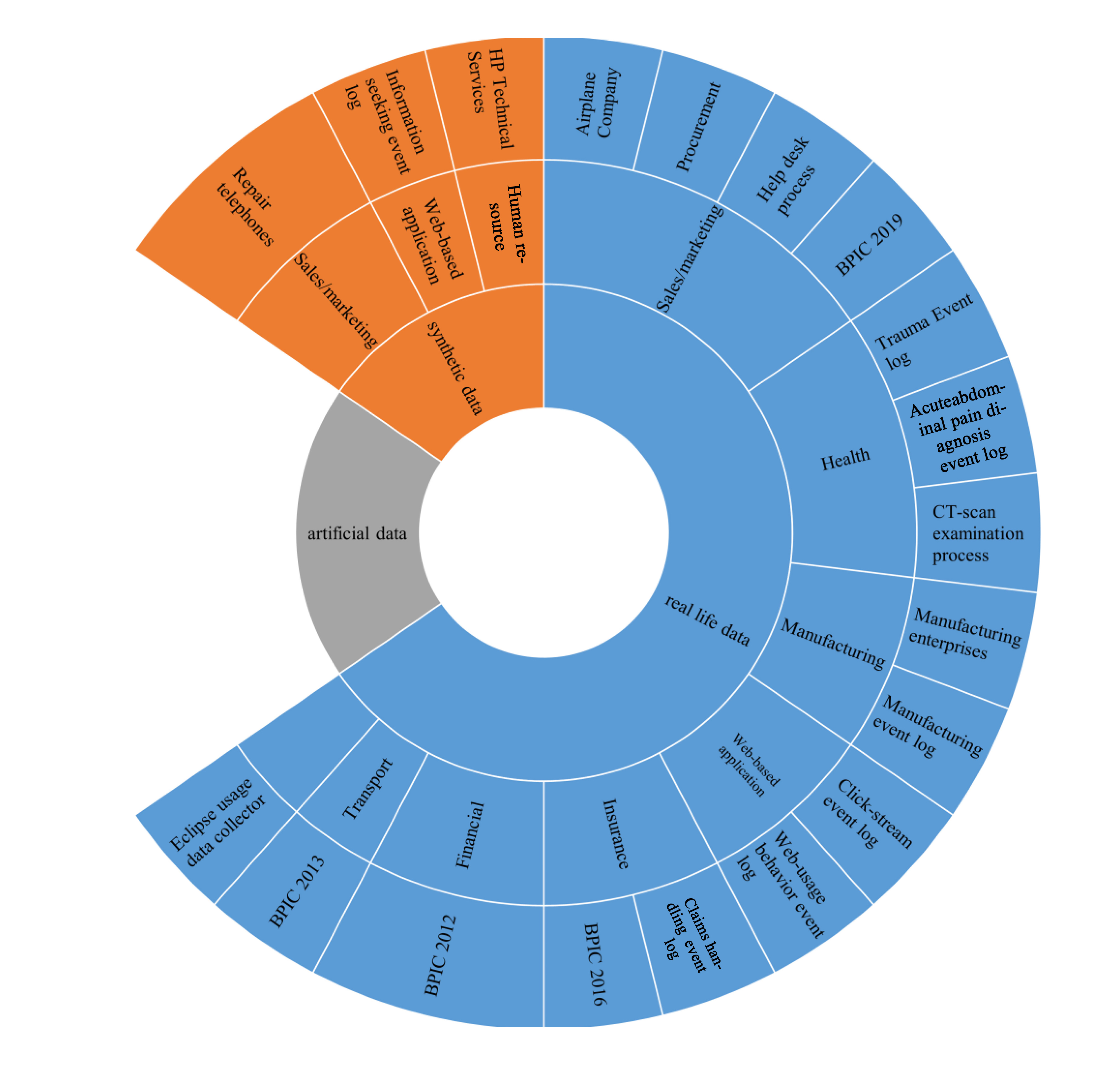}
\caption{The categories of methods applied per application domain }
\label{fig5}
\end{figure}

The details on application domains identified and the count of studies implemented for each are presented in \autoref{fig5}.  In this figure, the highest count of the studies belongs to sales/marketing \cite{Article_53,Article_56,Article_57,Article_65,Article_67}, web-based application \cite{Article_46,Article_54,Article_55} and healthcare \cite{Article_60,Article_61,Article_73}. Because PM is an emerging discipline and has not yet been applied in all fields, this section aims to reveal the  potential applicabilities in different fields and encourage researchers to explore these applications in the future.

\subsection{Implementation types}
\label{sec:14}
Many PM tools exist through which the event logs and other data generated by the processes are analyzed. About 30\% of methods are declared as being integrated as plugin or standalone tools on top of the ProM framework \cite{Article_39,Article_43,Article_44,Article_50,Article_51,Article_53,Article_58,Article_65,Article_67,Article_78}. ProM is a pluggable, open-source framework that allows the development of new PM algorithms.
 
The authors in \cite{Article_45} and \cite{Article_66} provide stand-alone plugin implementation on top of ConDec and YAWL\footnote{Yet Another Workflow Language}  , respectively. ConDec is a tool that proposes an open set of constraints for the high-level templates between BP activities and YAWL is a business process modeling language based on the workflow pattern and Petri nets. A plugin on top of CoCaMa, an ACM\footnote{Adaptive Case Management}  software solution, developed in \cite{Article_49}.

The method proposed by the authors in \cite{Article_54} is implemented as a stand-alone application using the Disco tool \cite{Article_80}, an easy-to-use process mining tool with a friendly interface design. The implementations of other articles are as a stand-alone tool in Java \cite{Article_60} and Eclipse environment \cite{Article_72,Article_78}, through Python \cite{Article_55,Article_56,Article_74} and in Matlab \cite{Article_52}.

Data analysis solutions have been applied mostly in NRR methods. Weka is applied in NRR cases \cite{Article_40,Article_61,Article_68,Article_70} and next to process analysis tools in NAR methods \cite{Article_66,Article_49}. RapidMiner is applied only in one NRR case \cite{Article_48}.

\begin{figure}
	
\includegraphics[width=1.1\textwidth]{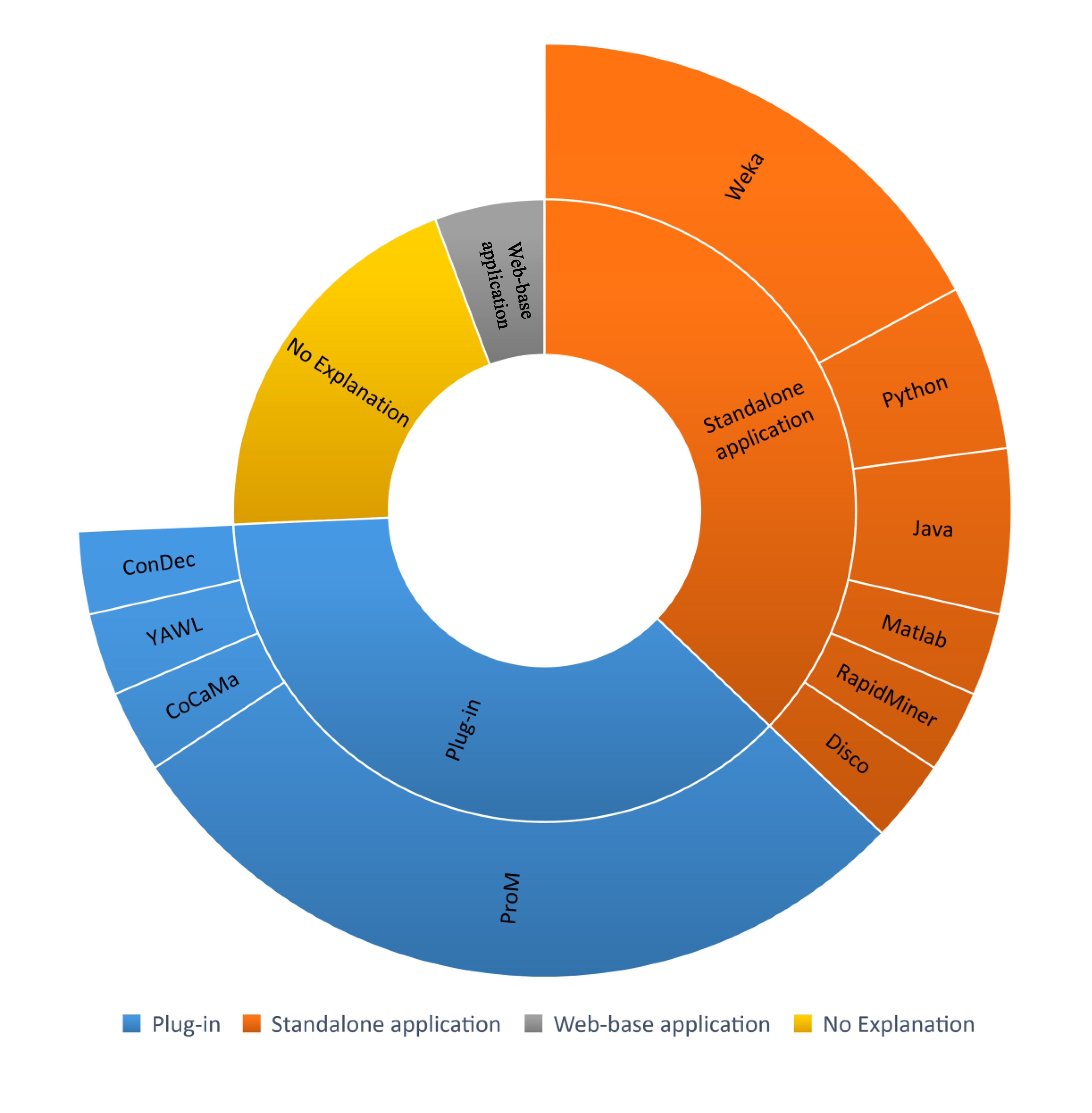}
\caption{Different implementation environments that are used in the PARS studies }
\label{fig6}
\end{figure}

In 20\% of studies run by \cite{Article_46,Article_54,Article_57,Article_69,Article_71,Article_73,Article_75}, there exists no declaration on any further explanation on implementation environments, next to two articles implemented as a web-based prototype without any further details \cite{Article_47,Article_59}. The implementation environments are shown in \autoref{fig6}. 

\subsection{Evaluation methods}
\label{sec:15}
Researchers in \cite{Article_81} sought to organize articles through classified research methods based on evaluation criteria in the three categories, i.e., \textit{proposal of the solution, validation research}, and \textit{evaluation research}. Validation research was applied to 60\% of primary studies to evaluate the proposed model. The assessment is run by doing experiments through simulation or synthetic data, prototyping, mathematical analysis, and mathematical proof of properties. It indicates an increase in the volume of studies with validation of the approaches through experiments where synthetic data are applied. 20\% of the primary studies have evaluation research, where the assessment of a proposal or implementation of a proposal is run by applying real scenarios and analyze the process owners' potential benefits. 20\% of the primary studies are classified according to the proposal of solution type. They propose a new solution technique without a full-blown validation accompanied by a simple example of a sound argument. It can be deduced that the proposed PARS approaches are often validated through prototyping, experiments, and applying synthetic data, with less tendency to the practical implementation of a solution regarding case studies. 

\subsection{Geographical analysis}
\label{sec:16}
\begin{figure}
	
\includegraphics[width=1\textwidth]{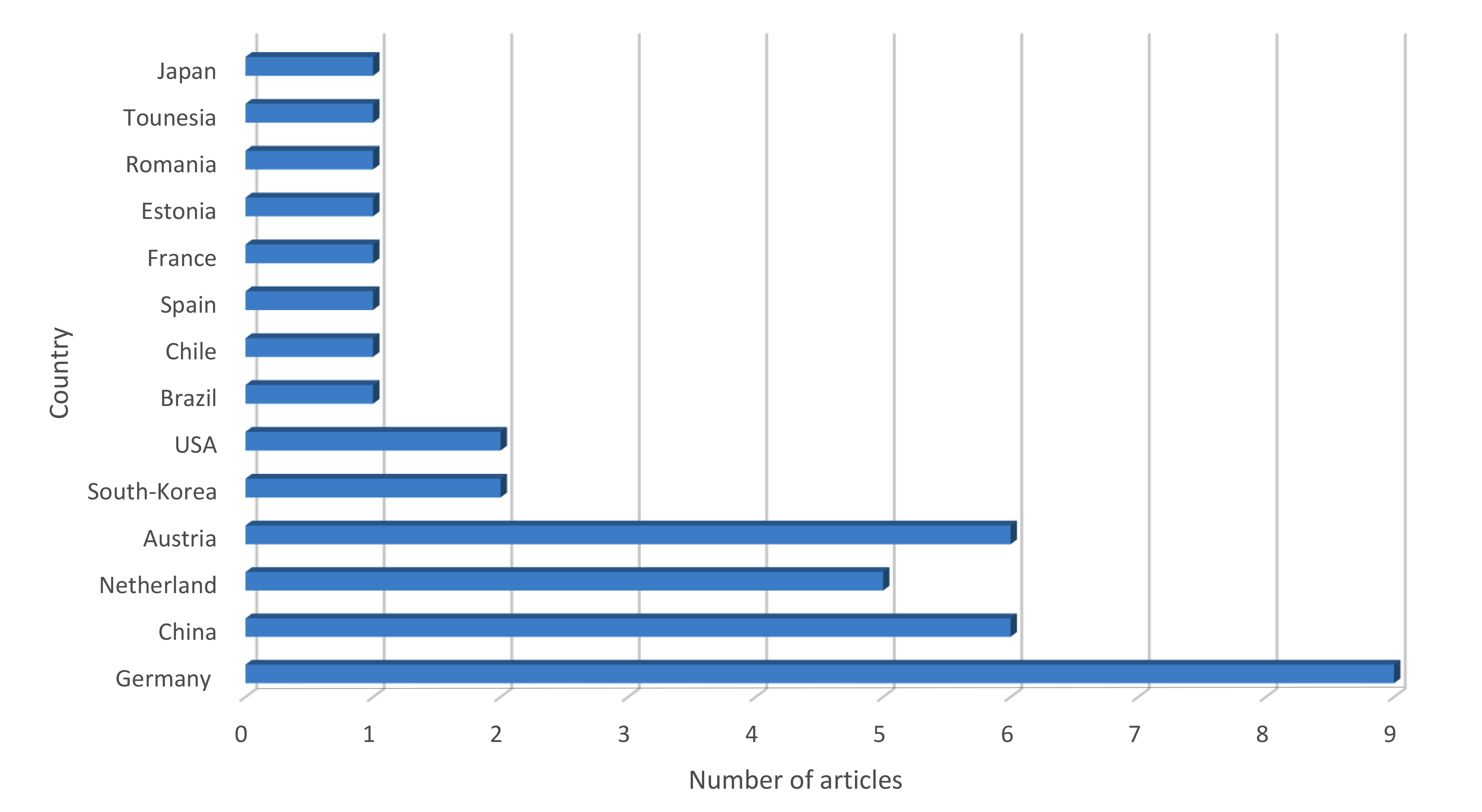}
\caption{Publication frequency by country}
\label{fig7}
\end{figure}

This type of analysis would contribute to determining the distribution of different research clusters at a global scale and identifying the international research groups dedicated to this particular research area. In general terms, the 33 studies were produced by 104 authors from 40 research institutions located in 15 countries. To run a detailed analysis on the premise and the authors' with the highest count of articles on process mining, only the most productive ones should be of concern.
\begin{itemize}
\item[-]	In 15 countries, the case studies of PARS approaches are evident. The countries with the highest publication frequency are ranked, as shown in \autoref{fig7}, are Germany, Austria, China, and the Netherlands. 
\item[-]	Among the research groups assessing the PARS research area, Department of Mathematics and Computer Science, Eindhoven University of Technology, the Netherlands, with five studies, the Software School of Fudan University, China and the Department of Computer Science, University of Innsbruck, Austria, with three studies each are outstanding.
\item[-]	Overall, 104 researchers have writing 33 studies, where Prof. van der Aalst with five studies, and Prof. Weber, Dr. de Leoni, and Prof. Weidong Zha with three studies each are the most active researchers in this field. 
\end{itemize}

The publications' geographical analysis corresponds to that of the top research groups in the PM field at a global scale. Next to the known researchers appeared through process mining-related publications also match with these groups. 

\section{Discussion}
\label{sec:17}
This article aims to integrate, analyze, and synthesize the proposals available in PARS research. By applying research questions in assessing the particular aspects of process-aware recommendation research, a classification framework is established to analyze 33 primary studies obtained through a systematic search. In the following, we first discuss some challenges in terms of PARSs, and afterward, some research gaps will be discussed. 

\subsection{Evaluation issues}
\label{sec:18}
Comparing performance in different approaches in publications makes it impossible to interpret and determine the most appropriate approach. According to \cite{Article_82}, the no free lunch theorem holds regarding search and optimization because no single optimization algorithm performs best in all situations. It illustrates that determining the most appropriate method depends on the customized data set, the application domain, different experimental settings, and the contextual factors, indicating that no comprehensive comparison of the performance values can be made among all studies. 

There exists no scientific measuring tool applicable in a direct comparison of the different PARSs. In this context, the count of distinguished proposals' empirical comparison in terms of performance is few. Selecting a proper assessment method highly depends on the objectives that a proposal concerning, the type of techniques and/or data applied for a recommendation. Though the existing PARS methods are objective-oriented, some studies have applied different event logs, experimental settings, assessment measures, and baselines, which has led to a situation with no possibility of exhaustive comparison in a methodologically and experimentally manner. This fact makes the researchers rely only on the potential optimistic impressions expressed in the respective articles. 

The lack of a clear framework for prescriptive monitoring techniques and/or universal measures to calculate the accuracy and validity of recommendations is another limitation recognized in this review study. It indicates that each assessment method's efficiency and applicability depend on the proposed technique therein. These issues reveal a degree of PARS studies' immaturity in implementing a standardized methodology for the assessment. Developing an integrated framework and applying more universal accuracy metrics applicable in all PARSs is a potential objective of future works.

In reviewing this literature, it is revealed that the details on experimentation are superficial and are not elaborated on the implementation environment. Moreover, the assessment methods are not explicitly stated in some studies. If this were not so, then the comparison therein would have been easier. Somehow, such studies have obtained good results, but there exists the doubt that due to author's bias, some results with low quality are not disclosed. For this purpose, some standard event logs are required for benchmarking, which will be proven to add value to the proposed methods. Only 40\% of the applied event logs are publicly available in this review study, 4TU ResearchData\footnote{https://data.4tu.nl} , and PODS4H\footnote{http://pods4h.com/pods4h} , allowing for the studies' reproducibility. About 75\% of this 40\% belong to real-life processes and make the validations of algorithms through real-life event logs more reliable and more substantial. Because artificial event logs do not necessarily reflect the complexity, variation, and unbalanced class distributions of real-life logs. Everything can be accomplished based on the synthetic data, while researchers must make sure that their simulated data reflect the actual process corresponding to the newly designed algorithm. Hence, we suggest evaluating algorithms on both synthetic and real-life event logs whenever possible. One of the important issues in future studies could be the focus on creating and sharing a collection of publicly available event logs that can be analyzed from multiple perspectives, especially in the operational support task. 

Furthermore, in almost all the studies, it is assumed that the system will work as it is after the recommendation. However, as shown in \cite{Article_74}, there is a possibility that the outcome of applying the recommended direction would not lead to the expected results. To overcome this challenge, utilizing the process-aware simulation methods at detail \cite{Article_83} and aggregated levels \cite{Article_84} besides PARS would be contributive. Using such approaches gives us this opportunity to evaluate the effect of the actions and measure the system's behavior after applying the recommendations.

\subsection{Size and cleanness of data }
\label{sec:19}
Increasing recorded data in information systems gives a higher potential to apply recommender systems. However, by increasing the size of data and the number of attributes, more hardware requirements and time are needed to provide a suitable PARS. In some scenarios, it is impractical to apply some of the existing methods using standard hardware. Moreover, in many real-life applications, data contain noise/outlier \cite{Article_85} and uncertainty \cite{Article_86}, leading to imprecise recommendations. 

The available PARSs are based on structured data for offline solutions, while the potential and challenges of such recommendations regarding big data scenarios with active process instances are yet to be assessed. Only a few works provide semi-online/real-time recommendations \cite{Article_72}. Applying and scaling PARSs on big data scenarios and stream processing is a promising research theme. These big data tools can present solutions to applied problems and introduce new inspiring challenges for further research. The developing models capable of handling complex data for assessing real-life problems will be developed. 

To overcome this challenge, one can adopt the current algorithms to work for big data scales and benefit from parallelization methods. Moreover, it is possible to reduce the size of data by wisely apply both domain-based and statistical pre-processing methods, e.g., sampling \cite{Article_2} and clustering \cite{Article_87}, before the recommendation to improve both the quality and performance of the final result. To handle active instances in a more efficient time, PARS can benefit from algorithms that work for streaming data \cite{Article_88,Article_89}.    

\subsection{Considering contextual information and frequent factors }
\label{sec:20}
In most realistic scenarios, numerous factors can affect the process, which should be considered in a PARS. A more challenging part is that different factors from different layers are available that affect each other in many of the cases. For example, the amount of workload can affect the speed of resources in the process such that up to a specific amount of tasks, resources perform better. Other factors such as new regulations or new advertisement methods can affect the number of new customers/users in the processes/systems. To deal with this challenge, one can propose using methods that can consider lots of parameters and their inter-connections \cite{Article_84}. Using this approach, different effective factors of the processes are extracted, and the relations between these factors and the external factors are modeled. 

The analyses here indicate that in most of the available studies, it is assumed that the process does not change in due time. Consequently, researchers do not estimate the impact of external factors and human behaviors in the process in a holistic perspective. It is observed that the contextual factors and different levels of abstraction are often omitted from prediction and recommendation aspects in PM. These proposed techniques have a common drawback, i.e., providing retrospective analyses and non-consideration of the environmental and external factors highly contributive in the process execution.
 
Though some studies have sought to consider contextual information, they tend to assess the trivial insight by only focusing on the process at a low-level of abstraction, thus, disabling in long-term predictions. Capturing different levels of abstractions and contextual information would generate more informative trustworthy recommendations. According to \cite{Article_6}, context information can improve the prediction quality as it adds valuable information to the predictive module. Consequently, analyzing just one model cannot produce a panorama on the relationships between resources and activities in a process. 

Analyzing the impact of some factors, like the effects of process changes, rich information hidden in human behavior, and dynamics in an ad-hoc environment in detail is a must in this area. The reason is that PARSs require extensive knowledge on business processes for long-term predictions considering multi-classes based on a context, outcome, or process variant. Recommendation(s)' quality is subject to the quality of the event logs and the discovered process model(s). The new techniques through which abstract views are generated on an organization in a process perspective based on instance data are considerably informative for top management if they provide performance insights. 

Customizing recommendations is an understudied essential theme. PARS must correspond to the users' profiles by learning their context and recognizing their objectives to generate recommendations proactively. Therefore, future research could deal with the proactivity and personalization challenges when developing recommender systems based on PM.
Furthermore, in most applications, providing completely automated recommender systems is inapplicable. For example, it is possible that some parts of a recommendation are not practical according to the business of the process. One way to deal with this challenge is providing interactive solutions and letting users apply their business knowledge interactively within the PARS. 

\subsection{Recommendation type}
\label{sec:21}
This study revealed that algorithms could be categorized according to the two main types of recommendation approaches of NAR and NRR. Among 33 articles, 19 are of the NAR category and 12 of NRR, and two are both. It is found that the path recommendation in the NAR category, a multi-resource recommendation in the NRR category, and recommendation in the multi-process instance environment of both categories are not assessed well yet. Process path recommendation of this type allows instance service organizations to improve customer satisfaction or detect unexpected termination(s). The importance of recommending multiple resources is important when a recommended resource is overloaded, and many instances are waiting in a queue for resource release. 

As to the recommendation strategies, three proposed models have focused on the process-aware recommendation issue through risk-minimizing; 14 proposals are related to metric-based and 11 proposals to pattern optimization. The risk-minimizing strategy is of less concern in these proposals. Risk-aware decision-making allows companies to detect and effectively manage the process instances to avoid the inappropriate effects of unexpected outcomes. When risk elimination is not feasible, process-related risk reduction and management become important.  

\subsection{Application domain and supporting framework }
\label{sec:22}
In this review study, the existence of tool support is of concern in determining whether it is a stand-alone application or a plugin in a broader platform. Such proposals with tool support help companies easily apply, assess, and understand the applicability, efficiency, and potential results of PARSs. 

A general multi-domain systematic study is conducted in \cite{Article_8}, to map the applications of PM. They identified 19 application domains, and for the top six areas, i.e., healthcare, ICT, manufacturing, education, finance, and logistics, they described the main contribution of PM. In this review study, we found that PARS approaches also have been applied in different domains. As stated in Subsection 4.4, sales/marketing, web-based application, and healthcare domains are the most active in this context. The algorithms for PARS are not limited and can be applied in new domains where the event log can be recorded. It would be of interest, especially for emerging smarter application domains where event logs are collected through sensors and contain behavioral and environmental data like the smart city, smart health, and smart grid. These domains are subject to new planning, execution, monitoring, and adapting process instances, because of the sensors, actuators, connectivity and analytics integration. 

In financial technology, the PARS techniques have potential application. The rapid growth and pervasive application of financial technologies have endangered the financial industry against different cyber-crime forms. Consequently, the financial services must resort to new advanced technologies therein to detect and prevent these fraudulent acts. This issue makes researchers re-assess the application of risk-minimizing PARS techniques in this industry. PARS techniques can be applied in predicting an individual's digital identity and detecting his/her fraudulent connections, and sound the alarm therefor. This new area of research seems to carry a broad set of promising unexplored aspects.

\subsection{Responsible PARS }
\label{sec:23}
With the advancement of data science techniques and the abundance of data, their applications and influences on human life expand profoundly. Their techniques are already applied in various fields, including life sciences, smart cities, transportation, and business intelligence. The huge influence of data science on human, make the hazards of data science without ethical considerations apparent. Disclosure of personal and confidential information and prejudice in the form of implicit bias in automated decision-making are two of the main risks of the irresponsible use of data science. These hazards have already been addressed and studied intensively in the area of data science, which results in methods and techniques that are compliant with ethical requirements by design. Moreover, in the area of PM, ethical issues have recently received attention.
 
Two of the most important aspects of responsibility in PM are fairness \cite{Article_90} and privacy \cite{Article_91}. In a fair PARS, the focus is on avoiding unfair recommendations (e.g., false or biased conclusions), even if they are accurate. In other words, For example, we should not consider the gender/religion/color of an applicant in a recommender system that works on top of a company's employment process. To consider privacy-preserving, a PARS should consider private and confidential data into account. According to plenty of regulations, such as GDPR\footnote{General Data Protection Regulation}  \cite{Article_92}, companies and information systems are obligated to consider individuals' privacy while analyzing their data. Therefore, it is necessary to consider which data attributes are stored by the PARS. Moreover, it should be considered which information can be revealed from the output of PARS that can conflict with privacy policies. These issues in PARSs are yet to be addressed.

\section{Threats to validity }
\label{sec:24}
According to the researchers in \cite{Article_93}, threats to validity are \textit{influences that may limit our ability to interpret or draw conclusions from the study's data}. They are of four main types: \textit{construct validity, internal validity, external validity,} and \textit{reliability} addressed here to be minimized in this literature review.

\begin{itemize}
\item[1-]\textit{Construct validity} focuses on the relation between the theory behind the experiment and the observation(s). It reflects to \textit{what extent the operational measures that are studied represent what the researcher has in mind and what is investigated according to the research questions}\cite{Article_94}. Accordingly, to minimize these threat types, combining the process mining and recommendation terms is well established to define a sufficiently stable search string. The count of articles found herein indicates that the search terms were well defined and reported. To reiterate, only available data between January 2008 and September 2020 are considered.
\item[2-] \textit{Internal validity} is related to the classification of each article according to the data collected by the authors. 
\item[3-] \textit{External validity} is concerned with to \textit{what extent it is possible to generalize the findings and the extent of their interest to other people outside the assessed case} \cite{Article_94}. In general, the obtained results are generalized and limited to the research period and the approaches published in the research area of recommendation systems and process mining disciplines. In this context, both the process mining and recommendation communities can benefit from this review's findings. In this regard, we included only published peer-reviewed journal/conference articles. 
\item[4-] \textit{Reliability} relates to data collection and the analysis thereof to gauge whether it was conducted in a way that others can repeat with the same results \cite{Article_95}. To maximize this goal, in this literature review, the guidelines outlined in \cite{Article_41} are applied, and the procedure is described in a level of detail that is sufficient to replicate the search. The search terms are defined according to a standard procedure.
\end{itemize}

One of this study's validity threats refers to the potential selection bias and inaccuracies in data extraction in the literature review. The well-known literature sources and libraries in information technology are applied to extract the relevant works on the PARS topic by running chain searching to avoid the exclusion of potentially relevant papers. 

Another potential bias is the subjectivity in applying the inclusion and exclusion criteria according to determining the subsumed studies. To mitigate this threat, each article is independently assessed against the inclusion and exclusion criteria by all the authors.

Another source of reliability threat is related to the count of returned records during the study period (i.e., from 2008 to 2020) because some articles can be registered in such databases only a few months after their publication. Consequently, some articles presented at the end of 2020 are not returned to our search pool. Researchers may produce different results as the search and ranking algorithms applied by source libraries are updated (see, e.g., Google Scholar). 

\section{Conclusion }
\label{sec:25}
This systematic review provides a solid foundation for future research on process-aware recommender systems. According to this survey, a limited number of studies exist where the PM improvements are applied for developing intelligence-based approach in making predictions about process outcomes, alarm generation for anomaly detection, and next activity/resource recommendation. During the past 30 years, PAIS research has focused on the core PAIS-relevant features' design like implementation, function, and application, with no concern on experience and feedback therein. It indicates that applying prescriptive analysis during the PAIS development has not been the main focus. To be able to use more beneficial analysis, the future research trend in PM should divert from pure descriptive analysis towards prescriptive analysis. 

No extensive systematic review study is conducted on the articles already published about PARS, which means the specific and new area between the PM and recommender systems' intersection. In this article, we have followed a protocol of conducting review studies to create an initial classification of the publications in this research area, something not considered in the available studies. This classification framework can be applied by researchers if obtaining a more comprehensive perspective on the theory and application of PARS. Such classification can support researchers involved in both the process mining and recommender systems to understand each other better to contribute to the development. This study answers the following questions:

\begin{itemize}
 \item[(1)]The trend of the published articles about PARS. The scope of this review covers 33 primary studies from 2008 to 2020 that have been published across journals (32\%), followed by conferences (58\%) and dissertations (10\%), next to other inclusion and exclusion criteria. During the 12 years, half of these studies have been disseminated in the past five years, which may indicate an upward trend in using PARS. Note that PARS is an active research area despite the prevailing limitations, and attempts should be made to keep pace with the ongoing advances made in this area. 
 \item[(2)]The common characteristics of the studies in which recommendation techniques are being applied to PM. As to the second research question, it was possible to delineate a picture of how the primary studies are distributed in terms of the type of recommendation methodologies, evaluation, geographical areas, and application domain aspects.
\end{itemize}

Regarding the second question, 55\% of studies belong to NAR, and 41\% of studies are metric-based. The most frequently applied process perspective is the control-flow (67\%), and few studies have considered contextual information (26\%). The publicly available event logs are applied in 40\% of studies, where only 30\% are developed based on the ProM framework. Evaluation methods of 56\% of studies are based on validation research. It indicates that approaches are not only validated using experiments that apply synthetic data and the implementation of prototypes but through the case studies as well. The increasing trend of case studies as an evaluation method promotes the maturity level of the research area. This greater maturity level, in turn, will require researchers in the future to assess their proposals in real environments. 

One of the main concerns in this review study is to find out to what extent the recommender systems algorithms and techniques would be applied in process mining, considering that both techniques are widely used in data science. To the authors here, in this context, although research in process mining is on rapid growth, little relevance is given to the recommendation problem, possibly due to a relative lack of knowledge about the potential benefits of such technique for process related-problems. An attempt is made to contribute to a change in this realm by publishing this review. The results obtained here confirm that the task of activity/resource recommendation is an emerging research area, and has attracted attentions over the last five years. 

According to this study's context, it is deduced that there is no applicable comprehensive process monitoring solution. A general model for the process-aware recommender system is yet to be proposed. An ideal PARS model should consider the contextual information and the impact of some factors, like the effects of changes in process, interpretability, and handling the great volumes of event logs in near-real-time. 
\newpage
\section*{Appendix: Details about event logs used in PARS}
\label{sec:26}
\begin{itemize}
 \item[-]	\textbf{BPI Challenge event logs}

\textbf{BPIC 2012 (Dutch financial institute event log):} The information of this real-life event log includes the process for a personal loan or overdraft in a global financing organization. This log was gathered from October 2011 to March 2012 and contains 13,087 cases and 262,200 events \cite{Article_67,Article_69}.

\textbf{BPIC 2013:} This event log is from Volvo IT Belgium and contains events from an incident and problem management system named the VINST. The data is separated into the three: 1- Incidents, a log of 7554 cases containing 65533 events related to the incident management process, 2- Closed problems, a log of 1487 cases containing 6660 events related to closed problems in the problem management process, and 3- Open problems, a log of 819 cases containing 2351 events related to open problems in the problem management process event logs \cite{Article_75}.
 
\textbf{BPIC 2016:} UWV is a Dutch governmental employee agency that provides financial support to Dutch residents who have lost their jobs and seek new employment. The data in this collection relates to the 300,000 applicants over a period of 8 months \cite{Article_74}.
 
\textbf{BPIC 2019: }This data is from a large multinational company operating in the coating and paint industry in Netherlands. There are collected over 1,5 million events of purchase orders submitted in 2018. The event log includes a total of 76,349 purchase documents containing a total of 251,734 items. Totally it contains 1,595,923 events relating to 42 activities performed by 627 users. Each case has the following attributes: name, purchasing document, item type, goods receipt, source, doc. category name, company, spend classification text, spend area text, sub spend area text, vendor, name, document type, item category \cite{Article_55}. 

 \item[-]\textbf{Other publicly available event logs}

\textbf{Eclipse Usage Data Collector (UDC): } is a system where the information on how do the developers apply the Eclipse platform is collected to generate event logs. The event logs are uploaded to servers hosted by the Eclipse foundation, which contains 1,048,576 event logs from over 10,000 Java developers between September 2009 and January 2010 \cite{Article_79}. 

\textbf{Repair telephone process:} This synthetic event log is related to a process in a telephone service repairing company, with 1104 cases and 11855 events. The minimum and maximum length of each trace is 4 and 25, respectively \cite{Article_67}. 

\textbf{Help desk process:} This dataset contains events from a ticketing management process of the help desk of an Italian software company. The process consists of nine activities: case ID, sub-process group, process typology, resource, cost, customer satisfaction, creation date, closing date, and priority. All cases begin with the insertion of a new ticket into the ticketing management system, and end when the issue is resolved. This event log contains 3804 process instances and 13710 events and it is available at https://data.mendeley.com/datasets/39bp3vv62t/1 \cite{Article_55,Article_65}. 

 \item[-]\textbf{Not publicly available event logs}

\textbf{HP technical services process:} This synthetic event log is gathered within an HP R\&D project on next generation IT service management solutions. More than 250 cases, related to issues in a human resource software application and incident management are collected. There are 4 types of cases, each including information like: case type, title, tags (a set of keywords that are attached to the case), step (activity), step flow (trace), participant, case status (draft, open, closed), priority (severity level of the issue) and contextual information on people interactions \cite{Article_72}.

\textbf{Information seeking process:} This synthetic data is applied in recommending optimized information seeking process to users and is divided into two separate parts: 1) User features, describing: keyword, web page, tag, access time, access date and location, and 2) Web page features describing: relation of tag/ tag category/ keyword and web page. This part contains information about seeking process start, seeking process end and successful seeking process. \cite{Article_46} 

\textbf{CT-scan examination process:} Radiology process collected from Chinese Huzhou Hospital's radiology department within 2009-06-01 to 2009-06-30 period. This real-life event log includes 244 cases and is not publically available because of patient privacy issues \cite{Article_61}.
 
\textbf{Claims handling process:} The not publicly available event log of a large insurance company recorded over one year containing 1065 traces \cite{Article_66}. 

\textbf{Manufacturing process:} This data is collected from a motor car shock absorber manufacturing factory with an event log containing 100,000 process instances\cite{Article_48}. 

\textbf{Acute abdominal pain diagnosis:} This real-life event log is of the experiments run within 2009–2010 and contains 11,490 cases. The total count of events is 80,430 where 82 employees participated in the process execution. The case ID, resources, cost, beginning and ending time are the data recorded for each task \cite{Article_73}.

\textbf{Airplane airline:} This event log contains a total of 2262 customer compensation data from a Shanghai airline. The data contains the count of customers for compensation, application time, compensation process operators, tickets ID for compensation, etc. \cite{Article_57}.

\textbf{Trauma process:} Three medical event logs are collected from the emergency department of children's national medical center, a level 1 pediatric trauma center in Washington, DC. Tracheal Intubation event log with ten context attributes: 1) Patient demographics (age, gender, BMI, injury type, injury severity score, pre-arrival intubation, mental status, body region injured), 2) Provider attributes (incubator’s medical role), 3) Event attributes (emergency/pre-arrival, direct-laryngoscopy/video-laryngoscopy and reason for incubation) \cite{Article_60}. 

\textbf{Click-stream process:} The data has click-stream information of 2 million users in a time frame of one month, with 10 million events \cite{Article_56}. 

\textbf{Procurement process:} This data is extracted from a SAP ERP system for procurement of goods in one month. It contains 33277cases, 255427 events, 37 items and 15 case attributes describing which good was ordered or which vendor delivered the order \cite{Article_53}. 

\textbf{Web-usage behavior process:} includes web usage behavior data of the customers from a financial service provider. The event log consists of a case ID, click ID, time stamp, and six context attributes. The data set includes 2,142 traces with a maximum length of 247 clicks \cite{Article_56}.

\textbf{Manufacturing enterprises:} This real-life event log is collected from three manufacturing enterprises, each containing three types of information: workflow history information, workflow model definition, and organizational information. Xiamen KingLong United Automotive Co. Ltd., contains 256 activities, 179 resources, and 10808 event entries; Hebei Zhongxing Automobile Co. Ltd., contains 399 activities, 244 resources and 42099 event entries and Datong Electronic Locomotive Co. Ltd. contains 922 activities, 147 resources and 99756 event entries \cite{Article_40}.
\end{itemize}

\bibliographystyle{spbasic}
          
	\bibliography{bibfile}

\begin{thebibliography}{95}
\providecommand{\natexlab}[1]{#1}
\providecommand{\url}[1]{{#1}}
\providecommand{\urlprefix}{URL }
\expandafter\ifx\csname urlstyle\endcsname\relax
  \providecommand{\doi}[1]{DOI~\discretionary{}{}{}#1}\else
  \providecommand{\doi}{DOI~\discretionary{}{}{}\begingroup
  \urlstyle{rm}\Url}\fi
\providecommand{\eprint}[2][]{\url{#2}}

\bibitem[{van~der Aalst(2008)}]{Article_5}
van~der Aalst WM (2008) Decision support based on process mining. In: Handbook
  on Decision Support Systems 1, Springer, pp 637--657

\bibitem[{van~der Aalst(2016)}]{Article_1}
van~der Aalst WMP (2016) Process Mining - Data Science in Action, Second
  Edition. Springer, \doi{10.1007/978-3-662-49851-4},
  \urlprefix\url{https://doi.org/10.1007/978-3-662-49851-4}

\bibitem[{van~der Aalst et~al.(2010)van~der Aalst, Pesic, and
  Song}]{Article_44}
van~der Aalst WMP, Pesic M, Song M (2010) Beyond process mining: From the past
  to present and future. In: Pernici B (ed) Advanced Information Systems
  Engineering, 22nd International Conference, CAiSE 2010, Hammamet, Tunisia,
  June 7-9, 2010. Proceedings, Springer, Lecture Notes in Computer Science, vol
  6051, pp 38--52, \doi{10.1007/978-3-642-13094-6\_5},
  \urlprefix\url{https://doi.org/10.1007/978-3-642-13094-6\_5}

\bibitem[{van~der Aalst et~al.(2016)van~der Aalst, Rosa, and
  Santoro}]{Article_34}
van~der Aalst WMP, Rosa ML, Santoro FM (2016) Business process management -
  don't forget to improve the process! Bus Inf Syst Eng 58(1):1--6,
  \doi{10.1007/s12599-015-0409-x},
  \urlprefix\url{https://doi.org/10.1007/s12599-015-0409-x}

\bibitem[{Abid et~al.(2016)Abid, Dominic, and Shahzad}]{Article_63}
Abid S, Dominic PDD, Shahzad K (2016) Business process analysis: a process
  warehouse-based resource preference evaluation method. Int J Bus Inf Syst
  21(2):137--161, \doi{10.1504/IJBIS.2016.074255},
  \urlprefix\url{https://doi.org/10.1504/IJBIS.2016.074255}

\bibitem[{Adam et~al.(2019)Adam, Alexandropoulos, Pardalos, and
  Vrahatis}]{Article_82}
Adam SP, Alexandropoulos SAN, Pardalos PM, Vrahatis MN (2019) No free lunch
  theorem: A review. Approximation and optimization pp 57--82

\bibitem[{Alves et~al.(2010)Alves, Niu, Alves, and Valen{\c{c}}a}]{Article_95}
Alves V, Niu N, Alves CF, Valen{\c{c}}a G (2010) Requirements engineering for
  software product lines: {A} systematic literature review. Inf Softw Technol
  52(8):806--820, \doi{10.1016/j.infsof.2010.03.014},
  \urlprefix\url{https://doi.org/10.1016/j.infsof.2010.03.014}

\bibitem[{Arias et~al.(2015)Arias, Rojas, Munoz{-}Gama, and
  Sep{\'{u}}lveda}]{Article_65}
Arias M, Rojas E, Munoz{-}Gama J, Sep{\'{u}}lveda M (2015) A framework for
  recommending resource allocation based on process mining. In: Reichert M,
  Reijers HA (eds) Business Process Management Workshops - {BPM} 2015, 13th
  International Workshops, Innsbruck, Austria, August 31 - September 3, 2015,
  Revised Papers, Springer, Lecture Notes in Business Information Processing,
  vol 256, pp 458--470, \doi{10.1007/978-3-319-42887-1\_37},
  \urlprefix\url{https://doi.org/10.1007/978-3-319-42887-1\_37}

\bibitem[{Arias et~al.(2018)Arias, Saavedra, Marques, Munoz-Gama, and
  Sep{\'u}lveda}]{Article_32}
Arias M, Saavedra R, Marques MR, Munoz-Gama J, Sep{\'u}lveda M (2018) Human
  resource allocation in business process management and process mining.
  Management Decision

\bibitem[{Augusto et~al.(2019)Augusto, Conforti, Dumas, Rosa, Maggi, Marrella,
  Mecella, and Soo}]{Article_23}
Augusto A, Conforti R, Dumas M, Rosa ML, Maggi FM, Marrella A, Mecella M, Soo A
  (2019) Automated discovery of process models from event logs: Review and
  benchmark. {IEEE} Trans Knowl Data Eng 31(4):686--705,
  \doi{10.1109/TKDE.2018.2841877},
  \urlprefix\url{https://doi.org/10.1109/TKDE.2018.2841877}

\bibitem[{Barba et~al.(2011)Barba, Weber, and Valle}]{Article_45}
Barba I, Weber B, Valle CD (2011) Supporting the optimized execution of
  business processes through recommendations. In: Daniel F, Barkaoui K, Dustdar
  S (eds) Business Process Management Workshops - {BPM} 2011 International
  Workshops, Clermont-Ferrand, France, August 29, 2011, Revised Selected
  Papers, Part {I}, Springer, Lecture Notes in Business Information Processing,
  vol~99, pp 135--140, \doi{10.1007/978-3-642-28108-2\_12},
  \urlprefix\url{https://doi.org/10.1007/978-3-642-28108-2\_12}

\bibitem[{Bogar{\'{\i}}n et~al.(2018)Bogar{\'{\i}}n, Cerezo, and
  Romero}]{Article_20}
Bogar{\'{\i}}n A, Cerezo R, Romero C (2018) A survey on educational process
  mining. Wiley Interdiscip Rev Data Min Knowl Discov 8(1),
  \doi{10.1002/widm.1230}, \urlprefix\url{https://doi.org/10.1002/widm.1230}

\bibitem[{Burattin(2019)}]{Article_89}
Burattin A (2019) Streaming process discovery and conformance checking. In:
  Sakr S, Zomaya AY (eds) Encyclopedia of Big Data Technologies, Springer,
  \doi{10.1007/978-3-319-63962-8\_103-1},
  \urlprefix\url{https://doi.org/10.1007/978-3-319-63962-8\_103-1}

\bibitem[{Chen et~al.(2013)Chen, Zhou, and Jin}]{Article_46}
Chen J, Zhou X, Jin Q (2013) Recommendation of optimized information seeking
  process based on the similarity of user access behavior patterns. Pers
  Ubiquitous Comput 17(8):1671--1681, \doi{10.1007/s00779-012-0601-7},
  \urlprefix\url{https://doi.org/10.1007/s00779-012-0601-7}

\bibitem[{Conforti et~al.(2015)Conforti, de~Leoni, Rosa, van~der Aalst, and ter
  Hofstede}]{Article_66}
Conforti R, de~Leoni M, Rosa ML, van~der Aalst WMP, ter Hofstede AHM (2015) A
  recommendation system for predicting risks across multiple business process
  instances. Decis Support Syst 69:1--19, \doi{10.1016/j.dss.2014.10.006},
  \urlprefix\url{https://doi.org/10.1016/j.dss.2014.10.006}

\bibitem[{Corallo et~al.(2020)Corallo, Lazoi, and Striani}]{Article_17}
Corallo A, Lazoi M, Striani F (2020) Process mining and industrial
  applications: A systematic literature review. Knowledge and Process
  Management 27(3):225--233

\bibitem[{Dakic et~al.(2018)Dakic, Stefanovic, Cosic, Lolic, and
  Medojevic}]{Article_11}
Dakic D, Stefanovic D, Cosic I, Lolic T, Medojevic M (2018) Business process
  mining application: A literature review. Annals of DAAAM \& Proceedings 29

\bibitem[{Dees et~al.(2019)Dees, de~Leoni, van~der Aalst, and
  Reijers}]{Article_74}
Dees M, de~Leoni M, van~der Aalst WMP, Reijers HA (2019) What if process
  predictions are not followed by good recommendations? In: vom Brocke J,
  Mendling J, Rosemann M (eds) Proceedings of the Industry Forum at {BPM} 2019
  co-located with 17th International Conference on Business Process Management
  {(BPM} 2019), Vienna, Austria, September 1-6, 2019, CEUR-WS.org, {CEUR}
  Workshop Proceedings, vol 2428, pp 61--72,
  \urlprefix\url{http://ceur-ws.org/Vol-2428/paper6.pdf}

\bibitem[{Detro et~al.(2020)Detro, Santos, Panetto, Loures, Lezoche, and
  Barra}]{Article_58}
Detro SP, Santos EAP, Panetto H, Loures EDFR, Lezoche M, Barra CMCM (2020)
  Applying process mining and semantic reasoning for process model
  customisation in healthcare. Enterp Inf Syst 14(7):983--1009,
  \doi{10.1080/17517575.2019.1632382},
  \urlprefix\url{https://doi.org/10.1080/17517575.2019.1632382}

\bibitem[{Dorn et~al.(2010)Dorn, Burkhart, Werth, and Dustdar}]{Article_47}
Dorn C, Burkhart T, Werth D, Dustdar S (2010) Self-adjusting recommendations
  for people-driven ad-hoc processes. In: Hull R, Mendling J, Tai S (eds)
  Business Process Management - 8th International Conference, {BPM} 2010,
  Hoboken, NJ, USA, September 13-16, 2010. Proceedings, Springer, Lecture Notes
  in Computer Science, vol 6336, pp 327--342,
  \doi{10.1007/978-3-642-15618-2\_23},
  \urlprefix\url{https://doi.org/10.1007/978-3-642-15618-2\_23}

\bibitem[{Dumas et~al.(2005)Dumas, van~der Aalst, and ter Hofstede}]{Article_3}
Dumas M, van~der Aalst WMP, ter Hofstede AHM (2005) Introduction. In: Dumas M,
  van~der Aalst WMP, ter Hofstede AHM (eds) Process-Aware Information Systems:
  Bridging People and Software Through Process Technology, Wiley, pp 1--20,
  \doi{10.1002/0471741442.ch1},
  \urlprefix\url{https://doi.org/10.1002/0471741442.ch1}

\bibitem[{Dunzer et~al.(2019)Dunzer, Stierle, Matzner, and Baier}]{Article_24}
Dunzer S, Stierle M, Matzner M, Baier S (2019) Conformance checking: a
  state-of-the-art literature review. In: Betz S (ed) Proceedings of the 11th
  International Conference on Subject-Oriented Business Process Management,
  {S-BPM} {ONE} 2019, Seville, Spain, June 26-28, 2019, {ACM}, pp 4:1--4:10,
  \doi{10.1145/3329007.3329014},
  \urlprefix\url{https://doi.org/10.1145/3329007.3329014}

\bibitem[{Erdogan and Tarhan(2018)}]{Article_12}
Erdogan T, Tarhan A (2018) Systematic mapping of process mining studies in
  healthcare. {IEEE} Access 6:24543--24567, \doi{10.1109/ACCESS.2018.2831244},
  \urlprefix\url{https://doi.org/10.1109/ACCESS.2018.2831244}

\bibitem[{Francescomarino et~al.(2018)Francescomarino, Ghidini, Maggi, and
  Milani}]{Article_29}
Francescomarino CD, Ghidini C, Maggi FM, Milani F (2018) Predictive process
  monitoring methods: Which one suits me best? In: Weske M, Montali M, Weber I,
  vom Brocke J (eds) Business Process Management - 16th International
  Conference, {BPM} 2018, Sydney, NSW, Australia, September 9-14, 2018,
  Proceedings, Springer, Lecture Notes in Computer Science, vol 11080, pp
  462--479, \doi{10.1007/978-3-319-98648-7\_27},
  \urlprefix\url{https://doi.org/10.1007/978-3-319-98648-7\_27}

\bibitem[{Ghasemi and Amyot(2016)}]{Article_14}
Ghasemi M, Amyot D (2016) Process mining in healthcare: a systematised
  literature review. Int J Electron Heal 9(1):60--88,
  \doi{10.1504/IJEH.2016.078745},
  \urlprefix\url{https://doi.org/10.1504/IJEH.2016.078745}

\bibitem[{Ghazal et~al.(2017)Ghazal, Ibrahim, and Salama}]{Article_21}
Ghazal MA, Ibrahim O, Salama MA (2017) Educational process mining: a systematic
  literature review. In: 2017 European Conference on Electrical Engineering and
  Computer Science (EECS), IEEE, pp 198--203

\bibitem[{Gr{\"{o}}ger et~al.(2014)Gr{\"{o}}ger, Schwarz, and
  Mitschang}]{Article_48}
Gr{\"{o}}ger C, Schwarz H, Mitschang B (2014) Prescriptive analytics for
  recommendation-based business process optimization. In: Abramowicz W,
  Kokkinaki AI (eds) Business Information Systems - 17th International
  Conference, {BIS} 2014, Larnaca, Cyprus, May 22-23, 2014. Proceedings,
  Springer, Lecture Notes in Business Information Processing, vol 176, pp
  25--37, \doi{10.1007/978-3-319-06695-0\_3},
  \urlprefix\url{https://doi.org/10.1007/978-3-319-06695-0\_3}

\bibitem[{Gr{\"{u}}ger et~al.(2020)Gr{\"{u}}ger, Bergmann, Kazik, and
  Kuhn}]{Article_13}
Gr{\"{u}}ger J, Bergmann R, Kazik Y, Kuhn M (2020) Process mining for case
  acquisition in oncology: {A} systematic literature review. In: Trabold D,
  Welke P, Piatkowski N (eds) Proceedings of the Conference "Lernen, Wissen,
  Daten, Analysen", Online, September 9-11, 2020, CEUR-WS.org, {CEUR} Workshop
  Proceedings, vol 2738, pp 162--173,
  \urlprefix\url{http://ceur-ws.org/Vol-2738/LWDA2020\_paper\_13.pdf}

\bibitem[{G{\"{u}}nther and Rozinat(2012)}]{Article_80}
G{\"{u}}nther CW, Rozinat A (2012) Disco: Discover your processes. In: Lohmann
  N, Moser S (eds) Proceedings of the Demonstration Track of the 10th
  International Conference on Business Process Management {(BPM} 2012),
  Tallinn, Estonia, September 4, 2012, CEUR-WS.org, {CEUR} Workshop
  Proceedings, vol 940, pp 40--44,
  \urlprefix\url{http://ceur-ws.org/Vol-940/paper8.pdf}

\bibitem[{Haisjackl and Weber(2010)}]{Article_43}
Haisjackl C, Weber B (2010) User assistance during process execution - an
  experimental evaluation of recommendation strategies. In: zur Muehlen M, Su J
  (eds) Business Process Management Workshops - {BPM} 2010 International
  Workshops and Education Track, Hoboken, NJ, USA, September 13-15, 2010,
  Revised Selected Papers, Springer, Lecture Notes in Business Information
  Processing, vol~66, pp 134--145, \doi{10.1007/978-3-642-20511-8\_12},
  \urlprefix\url{https://doi.org/10.1007/978-3-642-20511-8\_12}

\bibitem[{Huang et~al.(2011)Huang, Lu, and Duan}]{Article_61}
Huang Z, Lu X, Duan H (2011) Mining association rules to support resource
  allocation in business process management. Expert Syst Appl 38(8):9483--9490,
  \doi{10.1016/j.eswa.2011.01.146},
  \urlprefix\url{https://doi.org/10.1016/j.eswa.2011.01.146}

\bibitem[{Huber et~al.(2015)Huber, Fietta, and Hof}]{Article_49}
Huber S, Fietta M, Hof S (2015) Next step recommendation and prediction based
  on process mining in adaptive case management. In: Ehlers J, Thalheim B (eds)
  Proceedings of the 7th International Conference on Subject-Oriented Business
  Process Management, {S-BPM} {ONE} 2015, Kiel, Germany, April 23-24, 2015,
  {ACM}, pp 3:1--3:9, \doi{10.1145/2723839.2723842},
  \urlprefix\url{https://doi.org/10.1145/2723839.2723842}

\bibitem[{Jokonowo et~al.(2018)Jokonowo, Claes, Sarno, and
  Rochimah}]{Article_18}
Jokonowo B, Claes J, Sarno R, Rochimah S (2018) Process mining in supply
  chains: A systematic literature review. International Journal of Electrical
  \& Computer Engineering (2088-8708) 8(6)

\bibitem[{Khodabandelou(2013)}]{Article_78}
Khodabandelou G (2013) Contextual recommendations using intention mining on
  process traces: Doctoral consortium paper. In: Wieringa RJ, Nurcan S, Rolland
  C, Cavarero J (eds) {IEEE} 7th International Conference on Research
  Challenges in Information Science, {RCIS} 2013, Paris, France, May 29-31,
  2013, {IEEE}, pp 1--6, \doi{10.1109/RCIS.2013.6577728},
  \urlprefix\url{https://doi.org/10.1109/RCIS.2013.6577728}

\bibitem[{Khodabandelou(2014)}]{Article_79}
Khodabandelou G (2014) Mining intentional process models. PhD thesis,
  Universit{\'e} Panth{\'e}on-Sorbonne-Paris I

\bibitem[{Kitchenham(2012)}]{Article_42}
Kitchenham BA (2012) Systematic review in software engineering: where we are
  and where we should be going. In: Proceedings of the 2nd international
  workshop on Evidential assessment of software technologies, pp 1--2

\bibitem[{Kitchenham et~al.(2009)Kitchenham, Brereton, Budgen, Turner, Bailey,
  and Linkman}]{Article_41}
Kitchenham BA, Brereton P, Budgen D, Turner M, Bailey J, Linkman SG (2009)
  Systematic literature reviews in software engineering - {A} systematic
  literature review. Inf Softw Technol 51(1):7--15,
  \doi{10.1016/j.infsof.2008.09.009},
  \urlprefix\url{https://doi.org/10.1016/j.infsof.2008.09.009}

\bibitem[{Kluza et~al.(2013)Kluza, Baran, Bobek, and Nalepa}]{Article_31}
Kluza K, Baran M, Bobek S, Nalepa GJ (2013) Overview of recommendation
  techniques in business process modeling. In: Nalepa GJ, Baumeister J (eds)
  Proceedings of 9th Workshop on Knowledge Engineering and Software Engineering
  {(KESE9)} co-located with the 36th German Conference on Artificial
  Intelligence (KI2013), Koblenz, Germany, September 17, 2013, CEUR-WS.org,
  {CEUR} Workshop Proceedings, vol 1070,
  \urlprefix\url{http://ceur-ws.org/Vol-1070/kese9-05\_07.pdf}

\bibitem[{Koschmider et~al.(2011)Koschmider, Hornung, and
  Oberweis}]{Article_35}
Koschmider A, Hornung T, Oberweis A (2011) Recommendation-based editor for
  business process modeling. Data Knowl Eng 70(6):483--503,
  \doi{10.1016/j.datak.2011.02.002},
  \urlprefix\url{https://doi.org/10.1016/j.datak.2011.02.002}

\bibitem[{Lepenioti et~al.(2020)Lepenioti, Bousdekis, Apostolou, and
  Mentzas}]{Article_4}
Lepenioti K, Bousdekis A, Apostolou D, Mentzas G (2020) Prescriptive analytics:
  Literature review and research challenges. Int J Inf Manag 50:57--70,
  \doi{10.1016/j.ijinfomgt.2019.04.003},
  \urlprefix\url{https://doi.org/10.1016/j.ijinfomgt.2019.04.003}

\bibitem[{Li et~al.(2014)Li, Cao, Xu, Yin, Deng, Yin, and Wu}]{Article_38}
Li Y, Cao B, Xu L, Yin J, Deng S, Yin Y, Wu Z (2014) An efficient
  recommendation method for improving business process modeling. {IEEE} Trans
  Ind Informatics 10(1):502--513, \doi{10.1109/TII.2013.2258677},
  \urlprefix\url{https://doi.org/10.1109/TII.2013.2258677}

\bibitem[{Liu et~al.(2012)Liu, Cheng, and Ni}]{Article_68}
Liu T, Cheng Y, Ni Z (2012) Mining event logs to support workflow resource
  allocation. Knowl Based Syst 35:320--331, \doi{10.1016/j.knosys.2012.05.010},
  \urlprefix\url{https://doi.org/10.1016/j.knosys.2012.05.010}

\bibitem[{Liu et~al.(2008)Liu, Wang, Yang, and Sun}]{Article_40}
Liu Y, Wang J, Yang Y, Sun J (2008) A semi-automatic approach for workflow
  staff assignment. Comput Ind 59(5):463--476,
  \doi{10.1016/j.compind.2007.12.002},
  \urlprefix\url{https://doi.org/10.1016/j.compind.2007.12.002}

\bibitem[{Maamar et~al.(2016)Maamar, Faci, Sakr, Boukhebouze, and
  Barnawi}]{Article_62}
Maamar Z, Faci N, Sakr S, Boukhebouze M, Barnawi A (2016) Network-based social
  coordination of business processes. Inf Syst 58:56--74,
  \doi{10.1016/j.is.2016.02.005},
  \urlprefix\url{https://doi.org/10.1016/j.is.2016.02.005}

\bibitem[{Maggi et~al.(2014)Maggi, Di~Francescomarino, Dumas, and
  Ghidini}]{Article_28}
Maggi FM, Di~Francescomarino C, Dumas M, Ghidini C (2014) Predictive monitoring
  of business processes. In: International conference on advanced information
  systems engineering, Springer, pp 457--472

\bibitem[{Maita et~al.(2015)Maita, Martins, Paz, Peres, and
  Fantinato}]{Article_9}
Maita ARC, Martins LC, Paz CRL, Peres SM, Fantinato M (2015) Process mining
  through artificial neural networks and support vector machines: {A}
  systematic literature review. Bus Process Manag J 21(6):1391--1415,
  \doi{10.1108/BPMJ-02-2015-0017},
  \urlprefix\url{https://doi.org/10.1108/BPMJ-02-2015-0017}

\bibitem[{Maita et~al.(2018)Maita, Martins, Paz, Rafferty, Hung, Peres, and
  Fantinato}]{Article_7}
Maita ARC, Martins LC, Paz CRL, Rafferty L, Hung PCK, Peres SM, Fantinato M
  (2018) A systematic mapping study of process mining. Enterp Inf Syst
  12(5):505--549, \doi{10.1080/17517575.2017.1402371},
  \urlprefix\url{https://doi.org/10.1080/17517575.2017.1402371}

\bibitem[{M{\'{a}}rquez{-}Chamorro et~al.(2018)M{\'{a}}rquez{-}Chamorro,
  Resinas, and Ruiz{-}Cort{\'{e}}s}]{Article_6}
M{\'{a}}rquez{-}Chamorro AE, Resinas M, Ruiz{-}Cort{\'{e}}s A (2018) Predictive
  monitoring of business processes: {A} survey. {IEEE} Trans Serv Comput
  11(6):962--977, \doi{10.1109/TSC.2017.2772256},
  \urlprefix\url{https://doi.org/10.1109/TSC.2017.2772256}

\bibitem[{de~Medeiros et~al.(2007)de~Medeiros, Guzzo, Greco, van~der Aalst,
  Weijters, van Dongen, and Sacc{\`{a}}}]{Article_87}
de~Medeiros AKA, Guzzo A, Greco G, van~der Aalst WMP, Weijters AJMM, van Dongen
  BF, Sacc{\`{a}} D (2007) Process mining based on clustering: {A} quest for
  precision. In: ter Hofstede AHM, Benatallah B, Paik H (eds) Business Process
  Management Workshops, {BPM} 2007 International Workshops, BPI, BPD, CBP,
  ProHealth, RefMod, semantics4ws, Brisbane, Australia, September 24, 2007,
  Revised Selected Papers, Springer, Lecture Notes in Computer Science, vol
  4928, pp 17--29, \doi{10.1007/978-3-540-78238-4\_4},
  \urlprefix\url{https://doi.org/10.1007/978-3-540-78238-4\_4}

\bibitem[{Naderifar et~al.(2019)Naderifar, Sahran, and Shukur}]{Article_25}
Naderifar V, Sahran S, Shukur Z (2019) A review on conformance checking
  technique for the evaluation of process mining algorithms. TEM Journal
  8(4):1232

\bibitem[{Najar et~al.(2009)Najar, Saidani, Kirsch-Pinheiro, Souveyet, and
  Nurcan}]{Article_77}
Najar S, Saidani O, Kirsch-Pinheiro M, Souveyet C, Nurcan S (2009) Semantic
  representation of context models: a framework for analyzing and
  understanding. In: Proceedings of the 1st Workshop on Context, Information
  and Ontologies, pp 1--10

\bibitem[{Nakatumba et~al.(2012)Nakatumba, Westergaard, and van~der
  Aalst}]{Article_50}
Nakatumba J, Westergaard M, van~der Aalst WM (2012) A meta-model for
  operational support. BPM Center Report BPM-12-05, BPMcenter org pp 16--32

\bibitem[{Nezhad and Bartolini(2011)}]{Article_72}
Nezhad HRM, Bartolini C (2011) Next best step and expert recommendation for
  collaborative processes in {IT} service management. In: Rinderle{-}Ma S,
  Toumani F, Wolf K (eds) Business Process Management - 9th International
  Conference, {BPM} 2011, Clermont-Ferrand, France, August 30 - September 2,
  2011. Proceedings, Springer, Lecture Notes in Computer Science, vol 6896, pp
  50--61, \doi{10.1007/978-3-642-23059-2\_7},
  \urlprefix\url{https://doi.org/10.1007/978-3-642-23059-2\_7}

\bibitem[{Obregon et~al.(2013)Obregon, Kim, and Jung}]{Article_67}
Obregon J, Kim A, Jung J (2013) Dtminer: {A} tool for decision making based on
  historical process data. In: Song M, Wynn MT, Liu J (eds) Asia Pacific
  Business Process Management - First Asia Pacific Conference, {AP-BPM} 2013,
  Beijing, China, August 29-30, 2013. Selected Papers, Springer, Lecture Notes
  in Business Information Processing, vol 159, pp 81--91,
  \doi{10.1007/978-3-319-02922-1\_6},
  \urlprefix\url{https://doi.org/10.1007/978-3-319-02922-1\_6}

\bibitem[{Park and Song(2019)}]{Article_69}
Park G, Song M (2019) Prediction-based resource allocation using {LSTM} and
  minimum cost and maximum flow algorithm. In: International Conference on
  Process Mining, {ICPM} 2019, Aachen, Germany, June 24-26, 2019, {IEEE}, pp
  121--128, \doi{10.1109/ICPM.2019.00027},
  \urlprefix\url{https://doi.org/10.1109/ICPM.2019.00027}

\bibitem[{Pegoraro and van~der Aalst(2019)}]{Article_86}
Pegoraro M, van~der Aalst WMP (2019) Mining uncertain event data in process
  mining. In: International Conference on Process Mining, {ICPM} 2019, Aachen,
  Germany, June 24-26, 2019, {IEEE}, pp 89--96, \doi{10.1109/ICPM.2019.00023},
  \urlprefix\url{https://doi.org/10.1109/ICPM.2019.00023}

\bibitem[{Perry et~al.(2000)Perry, Porter, and Votta}]{Article_93}
Perry DE, Porter AA, Votta LG (2000) Empirical studies of software engineering:
  a roadmap. In: Finkelstein A (ed) 22nd International Conference on on
  Software Engineering, Future of Software Engineering Track, {ICSE} 2000,
  Limerick Ireland, June 4-11, 2000, {ACM}, pp 345--355,
  \doi{10.1145/336512.336586},
  \urlprefix\url{https://doi.org/10.1145/336512.336586}

\bibitem[{Petrusel and Stanciu(2012)}]{Article_51}
Petrusel R, Stanciu PL (2012) Making recommendations for decision processes
  based on aggregated decision data models. In: Abramowicz W, Kriksciuniene D,
  Sakalauskas V (eds) Business Information Systems - 15th International
  Conference, {BIS} 2012, Vilnius, Lithuania, May 21-23, 2012. Proceedings,
  Springer, Lecture Notes in Business Information Processing, vol 117, pp
  272--283, \doi{10.1007/978-3-642-30359-3\_24},
  \urlprefix\url{https://doi.org/10.1007/978-3-642-30359-3\_24}

\bibitem[{Polato et~al.(2018)Polato, Sperduti, Burattin, and
  de~Leoni}]{Article_33}
Polato M, Sperduti A, Burattin A, de~Leoni M (2018) Time and activity sequence
  prediction of business process instances. Computing 100(9):1005--1031,
  \doi{10.1007/s00607-018-0593-x},
  \urlprefix\url{https://doi.org/10.1007/s00607-018-0593-x}

\bibitem[{Pourbafrani et~al.(2019)Pourbafrani, van Zelst, and van~der
  Aalst}]{Article_84}
Pourbafrani M, van Zelst SJ, van~der Aalst WMP (2019) Scenario-based prediction
  of business processes using system dynamics. In: Panetto H, Debruyne C, Hepp
  M, Lewis D, Ardagna CA, Meersman R (eds) On the Move to Meaningful Internet
  Systems: {OTM} 2019 Conferences - Confederated International Conferences:
  CoopIS, ODBASE, C{\&}TC 2019, Rhodes, Greece, October 21-25, 2019,
  Proceedings, Springer, Lecture Notes in Computer Science, vol 11877, pp
  422--439, \doi{10.1007/978-3-030-33246-4\_27},
  \urlprefix\url{https://doi.org/10.1007/978-3-030-33246-4\_27}

\bibitem[{Qafari and van~der Aalst(2019)}]{Article_90}
Qafari MS, van~der Aalst WMP (2019) Fairness-aware process mining. In: Panetto
  H, Debruyne C, Hepp M, Lewis D, Ardagna CA, Meersman R (eds) On the Move to
  Meaningful Internet Systems: {OTM} 2019 Conferences - Confederated
  International Conferences: CoopIS, ODBASE, C{\&}TC 2019, Rhodes, Greece,
  October 21-25, 2019, Proceedings, Springer, Lecture Notes in Computer
  Science, vol 11877, pp 182--192, \doi{10.1007/978-3-030-33246-4\_11},
  \urlprefix\url{https://doi.org/10.1007/978-3-030-33246-4\_11}

\bibitem[{Rafiei and van~der Aalst(2020)}]{Article_91}
Rafiei M, van~der Aalst WMP (2020) Practical aspect of privacy-preserving data
  publishing in process mining. In: van~der Aalst WMP, vom Brocke J, Comuzzi M,
  Ciccio CD, Garc{\'{\i}}a F, Kumar A, Mendling J, Pentland BT, Pufahl L,
  Reichert M, Weske M (eds) Proceedings of the Best Dissertation Award,
  Doctoral Consortium, and Demonstration {\&} Resources Track at {BPM} 2020
  co-located with the 18th International Conference on Business Process
  Management {(BPM} 2020), Sevilla, Spain, September 13-18, 2020, CEUR-WS.org,
  {CEUR} Workshop Proceedings, vol 2673, pp 92--96,
  \urlprefix\url{http://ceur-ws.org/Vol-2673/paperDR06.pdf}

\bibitem[{Rama{-}Maneiro et~al.(2020)Rama{-}Maneiro, Vidal, and
  Lama}]{Article_26}
Rama{-}Maneiro E, Vidal JC, Lama M (2020) Deep learning for predictive business
  process monitoring: Review and benchmark. CoRR abs/2009.13251,
  \urlprefix\url{https://arxiv.org/abs/2009.13251}, \eprint{2009.13251}

\bibitem[{Rivas and Bayona-Or{\'e}(2019)}]{Article_22}
Rivas MH, Bayona-Or{\'e} S (2019) Process mining algorithms for automated
  process discovery. RISTI-Revista Iberica de Sistemas e Tecnologias de
  Informacao pp 33--49

\bibitem[{Rojas et~al.(2016)Rojas, Munoz{-}Gama, Sep{\'{u}}lveda, and
  Capurro}]{Article_15}
Rojas E, Munoz{-}Gama J, Sep{\'{u}}lveda M, Capurro D (2016) Process mining in
  healthcare: {A} literature review. J Biomed Informatics 61:224--236,
  \doi{10.1016/j.jbi.2016.04.007},
  \urlprefix\url{https://doi.org/10.1016/j.jbi.2016.04.007}

\bibitem[{Rozinat et~al.(2009)Rozinat, Mans, Song, and van~der
  Aalst}]{Article_83}
Rozinat A, Mans RS, Song M, van~der Aalst WMP (2009) Discovering simulation
  models. Inf Syst 34(3):305--327, \doi{10.1016/j.is.2008.09.002},
  \urlprefix\url{https://doi.org/10.1016/j.is.2008.09.002}

\bibitem[{Sani et~al.(2017)Sani, van Zelst, and van~der Aalst}]{Article_85}
Sani MF, van Zelst SJ, van~der Aalst WMP (2017) Improving process discovery
  results by filtering outliers using conditional behavioural probabilities.
  In: Teniente E, Weidlich M (eds) Business Process Management Workshops -
  {BPM} 2017 International Workshops, Barcelona, Spain, September 10-11, 2017,
  Revised Papers, Springer, Lecture Notes in Business Information Processing,
  vol 308, pp 216--229, \doi{10.1007/978-3-319-74030-0\_16},
  \urlprefix\url{https://doi.org/10.1007/978-3-319-74030-0\_16}

\bibitem[{Sani et~al.(2021)Sani, van Zelst, and van~der Aalst}]{Article_2}
Sani MF, van Zelst SJ, van~der Aalst WM (2021) The impact of biased sampling of
  event logs on the performance of process discovery. Computing pp 1--20

\bibitem[{dos Santos~Garcia et~al.(2019)dos Santos~Garcia, Meincheim, Junior,
  Dallagassa, Sato, Carvalho, Santos, and Scalabrin}]{Article_8}
dos Santos~Garcia C, Meincheim A, Junior ERF, Dallagassa MR, Sato DMV, Carvalho
  DR, Santos EAP, Scalabrin EE (2019) Process mining techniques and
  applications - {A} systematic mapping study. Expert Syst Appl 133:260--295,
  \doi{10.1016/j.eswa.2019.05.003},
  \urlprefix\url{https://doi.org/10.1016/j.eswa.2019.05.003}

\bibitem[{Schobel and Reichert(2017)}]{Article_52}
Schobel J, Reichert M (2017) A predictive approach enabling process execution
  recommendations. In: Grambow G, Oberhauser R, Reichert M (eds) Advances in
  Intelligent Process-Aware Information Systems - Concepts, Methods, and
  Technologies, Intelligent Systems Reference Library, vol 123, Springer
  International Publishing, pp 155--170, \doi{10.1007/978-3-319-52181-7\_6},
  \urlprefix\url{https://doi.org/10.1007/978-3-319-52181-7\_6}

\bibitem[{Schonenberg et~al.(2008)Schonenberg, Weber, van Dongen, and van~der
  Aalst}]{Article_39}
Schonenberg H, Weber B, van Dongen BF, van~der Aalst WMP (2008) Supporting
  flexible processes through recommendations based on history. In: Dumas M,
  Reichert M, Shan M (eds) Business Process Management, 6th International
  Conference, {BPM} 2008, Milan, Italy, September 2-4, 2008. Proceedings,
  Springer, Lecture Notes in Computer Science, vol 5240, pp 51--66,
  \doi{10.1007/978-3-540-85758-7\_7},
  \urlprefix\url{https://doi.org/10.1007/978-3-540-85758-7\_7}

\bibitem[{Sch{\"{o}}nig et~al.(2012)Sch{\"{o}}nig, Zeising, and
  Jablonski}]{Article_70}
Sch{\"{o}}nig S, Zeising M, Jablonski S (2012) Adapting association rule mining
  to discover patterns of collaboration in process logs. In: Pu C, Joshi J,
  Nepal S (eds) 8th International Conference on Collaborative Computing:
  Networking, Applications and Worksharing, CollaborateCom 2012, Pittsburgh,
  PA, USA, October 14-17, 2012, {ICST} / {IEEE}, pp 531--534,
  \doi{10.4108/icst.collaboratecom.2012.250346},
  \urlprefix\url{https://doi.org/10.4108/icst.collaboratecom.2012.250346}

\bibitem[{Seeliger et~al.(2018)Seeliger, Nolle, and
  M{\"u}hlh{\"a}user}]{Article_53}
Seeliger A, Nolle T, M{\"u}hlh{\"a}user M (2018) Processexplorer: An
  interactive visual recommendation system for process mining. In: KDD Workshop
  on Interactive Data Exploration and Analytics

\bibitem[{Setiawan et~al.(2011)Setiawan, Sadiq, and Kirkman}]{Article_36}
Setiawan MA, Sadiq SW, Kirkman R (2011) Facilitating business process
  improvement through personalized recommendation. In: Abramowicz W (ed)
  Business Information Systems - 14th International Conference, {BIS} 2011,
  Poznan, Poland, June 15-17, 2011. Proceedings, Springer, Lecture Notes in
  Business Information Processing, vol~87, pp 136--147,
  \doi{10.1007/978-3-642-21863-7\_12},
  \urlprefix\url{https://doi.org/10.1007/978-3-642-21863-7\_12}

\bibitem[{Sindhgatta et~al.(2016)Sindhgatta, Ghose, and Dam}]{Article_75}
Sindhgatta R, Ghose AK, Dam HK (2016) Context-aware analysis of past process
  executions to aid resource allocation decisions. In: Nurcan S, Soffer P,
  Bajec M, Eder J (eds) Advanced Information Systems Engineering - 28th
  International Conference, CAiSE 2016, Ljubljana, Slovenia, June 13-17, 2016.
  Proceedings, Springer, Lecture Notes in Computer Science, vol 9694, pp
  575--589, \doi{10.1007/978-3-319-39696-5\_35},
  \urlprefix\url{https://doi.org/10.1007/978-3-319-39696-5\_35}

\bibitem[{Teinemaa et~al.(2019)Teinemaa, Dumas, Rosa, and Maggi}]{Article_27}
Teinemaa I, Dumas M, Rosa ML, Maggi FM (2019) Outcome-oriented predictive
  process monitoring: Review and benchmark. {ACM} Trans Knowl Discov Data
  13(2):17:1--17:57, \doi{10.1145/3301300},
  \urlprefix\url{https://doi.org/10.1145/3301300}

\bibitem[{Terragni and Hassani(2018)}]{Article_54}
Terragni A, Hassani M (2018) Analyzing customer journey with process mining:
  From discovery to recommendations. In: Younas M, Disso JP (eds) 6th {IEEE}
  International Conference on Future Internet of Things and Cloud, FiCloud
  2018, Barcelona, Spain, August 6-8, 2018, {IEEE} Computer Society, pp
  224--229, \doi{10.1109/FiCloud.2018.00040},
  \urlprefix\url{https://doi.org/10.1109/FiCloud.2018.00040}

\bibitem[{Thiede et~al.(2018)Thiede, Fuerstenau, and Barquet}]{Article_19}
Thiede M, Fuerstenau D, Barquet APB (2018) How is process mining technology
  used by organizations? {A} systematic literature review of empirical studies.
  Bus Process Manag J 24(4):900--922, \doi{10.1108/BPMJ-06-2017-0148},
  \urlprefix\url{https://doi.org/10.1108/BPMJ-06-2017-0148}

\bibitem[{Triki et~al.(2013)Triki, Saoud, Dugdale, and Hanachi}]{Article_59}
Triki S, Saoud NBB, Dugdale J, Hanachi C (2013) Coupling case based reasoning
  and process mining for a web based crisis management decision support system.
  In: Reddy S, Jmaiel M (eds) 2013 Workshops on Enabling Technologies:
  Infrastructure for Collaborative Enterprises, Hammamet, Tunisia, June 17-20,
  2013, {IEEE} Computer Society, pp 245--252, \doi{10.1109/WETICE.2013.77},
  \urlprefix\url{https://doi.org/10.1109/WETICE.2013.77}

\bibitem[{Van Der~Aalst and Dustdar(2012)}]{Article_76}
Van Der~Aalst WM, Dustdar S (2012) Process mining put into context. IEEE
  Internet Computing 16(1):82--86

\bibitem[{Verenich et~al.(2019)Verenich, Dumas, Rosa, Maggi, and
  Teinemaa}]{Article_30}
Verenich I, Dumas M, Rosa ML, Maggi FM, Teinemaa I (2019) Survey and
  cross-benchmark comparison of remaining time prediction methods in business
  process monitoring. ACM Transactions on Intelligent Systems and Technology
  (TIST) 10(4):1--34

\bibitem[{Voss(2016)}]{Article_92}
Voss WG (2016) European union data privacy law reform: General data protection
  regulation, privacy shield, and the right to delisting. The Business Lawyer
  72(1):221--234

\bibitem[{Weinzierl et~al.(2020{\natexlab{a}})Weinzierl, Dunzer, Zilker, and
  Matzner}]{Article_55}
Weinzierl S, Dunzer S, Zilker S, Matzner M (2020{\natexlab{a}}) Prescriptive
  business process monitoring for recommending next best actions. In: Fahland
  D, Ghidini C, Becker J, Dumas M (eds) Business Process Management Forum -
  {BPM} Forum 2020, Seville, Spain, September 13-18, 2020, Proceedings,
  Springer, Lecture Notes in Business Information Processing, vol 392, pp
  193--209, \doi{10.1007/978-3-030-58638-6\_12},
  \urlprefix\url{https://doi.org/10.1007/978-3-030-58638-6\_12}

\bibitem[{Weinzierl et~al.(2020{\natexlab{b}})Weinzierl, Stierle, Zilker, and
  Matzner}]{Article_56}
Weinzierl S, Stierle M, Zilker S, Matzner M (2020{\natexlab{b}}) A next click
  recommender system for web-based service analytics with context-aware lstms.
  In: 53rd Hawaii International Conference on System Sciences, {HICSS} 2020,
  Maui, Hawaii, USA, January 7-10, 2020, ScholarSpace, pp 1--10,
  \urlprefix\url{http://hdl.handle.net/10125/63929}

\bibitem[{Wibisono et~al.(2015)Wibisono, Nisafani, Bae, and Park}]{Article_64}
Wibisono A, Nisafani AS, Bae H, Park Y (2015) On-the-fly performance-aware
  human resource allocation in the business process management systems
  environment using na{\"{\i}}ve bayes. In: Bae J, Suriadi S, Wen L (eds) Asia
  Pacific Business Process Management - Third Asia Pacific Conference, {AP-BPM}
  2015, Busan, South Korea, June 24-26, 2015, Proceedings, Springer, Lecture
  Notes in Business Information Processing, vol 219, pp 70--80,
  \doi{10.1007/978-3-319-19509-4\_6},
  \urlprefix\url{https://doi.org/10.1007/978-3-319-19509-4\_6}

\bibitem[{Wieringa et~al.(2006)Wieringa, Maiden, Mead, and
  Rolland}]{Article_81}
Wieringa RJ, Maiden NAM, Mead NR, Rolland C (2006) Requirements engineering
  paper classification and evaluation criteria: a proposal and a discussion.
  Requir Eng 11(1):102--107, \doi{10.1007/s00766-005-0021-6},
  \urlprefix\url{https://doi.org/10.1007/s00766-005-0021-6}

\bibitem[{Wohlin et~al.(2012)Wohlin, Runeson, H{\"{o}}st, Ohlsson, and
  Regnell}]{Article_94}
Wohlin C, Runeson P, H{\"{o}}st M, Ohlsson MC, Regnell B (2012) Experimentation
  in Software Engineering. Springer, \doi{10.1007/978-3-642-29044-2},
  \urlprefix\url{https://doi.org/10.1007/978-3-642-29044-2}

\bibitem[{Yang et~al.(2017)Yang, Dong, Sun, Zhou, Farneth, Xiong, Burd, and
  Marsic}]{Article_60}
Yang S, Dong X, Sun L, Zhou Y, Farneth RA, Xiong H, Burd RS, Marsic I (2017) A
  data-driven process recommender framework. In: Proceedings of the 23rd {ACM}
  {SIGKDD} International Conference on Knowledge Discovery and Data Mining,
  Halifax, NS, Canada, August 13 - 17, 2017, {ACM}, pp 2111--2120,
  \doi{10.1145/3097983.3098174},
  \urlprefix\url{https://doi.org/10.1145/3097983.3098174}

\bibitem[{Yang and Su(2014)}]{Article_16}
Yang W, Su Q (2014) Process mining for clinical pathway: Literature review and
  future directions. In: 2014 11th International Conference on Service Systems
  and Service Management (ICSSSM), IEEE, pp 1--5

\bibitem[{Yue et~al.(2011)Yue, Wu, Wang, and Bai}]{Article_10}
Yue D, Wu X, Wang H, Bai J (2011) A review of process mining algorithms. In:
  2011 International Conference on Business Management and Electronic
  Information, IEEE, vol~5, pp 181--185

\bibitem[{van Zelst(2019)}]{Article_88}
van Zelst S (2019) Process mining with streaming data. Technische Universiteit
  Eindhoven

\bibitem[{Zhao et~al.(2015)Zhao, Yang, Liu, and Wu}]{Article_71}
Zhao W, Yang L, Liu H, Wu R (2015) The optimization of resource allocation
  based on process mining. In: Huang D, Han K (eds) Advanced Intelligent
  Computing Theories and Applications - 11th International Conference, {ICIC}
  2015, Fuzhou, China, August 20-23, 2015. Proceedings, Part {III}, Springer,
  Lecture Notes in Computer Science, vol 9227, pp 341--353,
  \doi{10.1007/978-3-319-22053-6\_38},
  \urlprefix\url{https://doi.org/10.1007/978-3-319-22053-6\_38}

\bibitem[{Zhao et~al.(2016)Zhao, Liu, Dai, and Ma}]{Article_73}
Zhao W, Liu H, Dai W, Ma J (2016) An entropy-based clustering ensemble method
  to support resource allocation in business process management. Knowl Inf Syst
  48(2):305--330, \doi{10.1007/s10115-015-0879-7},
  \urlprefix\url{https://doi.org/10.1007/s10115-015-0879-7}

\bibitem[{Zhao et~al.(2017)Zhao, Zeng, Zheng, and Yang}]{Article_57}
Zhao W, Zeng Q, Zheng G, Yang L (2017) The resource allocation model for
  multi-process instances based on particle swarm optimization. Inf Syst
  Frontiers 19(5):1057--1066, \doi{10.1007/s10796-017-9743-5},
  \urlprefix\url{https://doi.org/10.1007/s10796-017-9743-5}

\bibitem[{Zhen et~al.(2009)Zhen, Huang, and Jiang}]{Article_37}
Zhen L, Huang GQ, Jiang Z (2009) Recommender system based on workflow. Decis
  Support Syst 48(1):237--245, \doi{10.1016/j.dss.2009.08.002},
  \urlprefix\url{https://doi.org/10.1016/j.dss.2009.08.002}

\end{thebibliography}

\end{document}